\documentclass[final,3p,times,twocolumn]{elsarticle}

\usepackage{amssymb,amsmath,amsthm,bm}
\usepackage{geometry}
\usepackage{algorithm}
\usepackage{algpseudocode}
\usepackage{empheq}
\usepackage{graphicx}
\graphicspath{{figures-pdf//}}
\usepackage[normalem]{ulem}
\usepackage{url}
\usepackage{cases}
\usepackage{subeqnarray}
\usepackage{slashbox}
\usepackage{bm}
\usepackage{graphicx,color,overpic}
\usepackage{float}
\usepackage{booktabs}

\newlength\figsep
\usepackage{enumitem}
\newlength\OneImW
\newlength\TwoImW
\newlength\BigOneImW
\newlength\ThreeImW
\newlength\DoubleThreeImW
\newlength\sfigwidth
\journal{IEEE Multimedia}
\usepackage{lineno}

\usepackage[font=small,skip=0pt]{caption}
\usepackage[bookmarks=false]{hyperref}
\hypersetup{
 linktocpage=true,
 pdfborderstyle={/S/S/W 1},
 hyperindex=true,
 bookmarks=false,
 bookmarksopen=false,
 bookmarksnumbered=false,
}

\begin{document}

\begin{frontmatter}

\title{Cryptanalyzing two image encryption algorithms based on a first-order time-delay system}

\author[hnu-cn]{Sheng Liu}

\author[xtu-cn]{Chengqing Li\corref{corr}}
\ead{DrChengqingLi@gmail.com}
\cortext[corr]{Corresponding author.}

\author[hnu-cn]{Qiao Hu}

\address[hnu-cn]{School of Computer Science and Electronic Engineering, Hunan University, Changsha 410082, Hunan, China}

\address[xtu-cn]{School of Computer Science, Xiangtan University, Xiangtan 411105, Hunan, China}

\begin{abstract}
Security is a key problem for the transmission, interchange and storage process of multimedia systems and applications.
In 2018, M. Li et al. proposed in-depth security analysis on an image encryption algorithm based on a first-order time-delay system (IEATD) and gave a specific chosen-plaintext attack on it. Moreover, an enhanced version called as IEACD was designed to fix the reported security defects.
This paper analyzes the essential structures of the two algorithms and evaluates their real security performances:
1) no efficient nonlinear operations are adopted to assure the sensibility of keystream;
2) the equivalent secret key of IEATD can be efficiently recovered from one known plain-image and the corresponding cipher-image;
3) IEACD can still be efficiently cracked with a chosen-plaintext attack.
Both rigorous theoretical analyses and detailed experimental results are provided to demonstrate effectiveness of the advanced cryptanalytic methods.
\end{abstract}

\end{frontmatter}

\section{Introduction}

Social media not only drive product discovery and purchase, but also incur serious concern on the security and privacy of the images shared
by the Internet users. Due to the special properties of multimedia information, the modern text encryption standards, such as AES and Triple DES, cannot efficiently protect them in general.
To cope with the challenge, a number of special image encryption algorithms, e.g. joint encryption and compression together, were proposed every year \cite{abu:CA:IM12,yegd:auto:IM16,cqli:meet:JISA19}.
It is well known that cryptography (designing encryption algorithm) and cryptanalysis (security analysis of a given encryption algorithm) are two integral parts of cryptology.
The cryptanalysis results facilitate the designers strengthen or replace flawed algorithms. Cryptanalysis of a given image encryption scheme also provides a special perspective for promoting some multimedia processing techniques, e.g. image recovery.
Some image encryption algorithms like that proposed in \cite{Mannai:delay:ND15,yegd:auto:IM16} are found to be insecure to different extents from the viewpoint of modern cryptology
\cite{Jolfaei:Permut:TIFS16,cqli:autoblock:IEEEM18,Preishuber:motivation:TIFS2018,Chenjx:psn:TCSVT21}.

The complex dynamics of a chaotic system demonstrated in an infinite-precision domain is very similar to the expected properties of a secure
encryption scheme outlined by Shannon in \cite{Shannon:CommunicationTheory1949}. So, a large number of chaos-based encryption schemes were proposed in the past
three decades \cite{chai:DNA:SP19,Hua2021ND}. In \cite{ikeda:1980:optical}, Ikeda adopted a one-variable differential-difference equation to model light going around a ring cavity containing a nonlinear dielectric medium and found ``chaotic" phenomena in the transmitted field.
In \cite{Mannai:delay:ND15}, Mannai et al. introduced the equation's variant
\begin{equation}
\label{eq:delay}
\frac{dx(t)}{dt} = -\alpha x(t) + m \sin(x(t-T))
\end{equation}
as a chaos-based pseudorandom number generator (PRNG), where $\alpha$ and $m$ are coefficients, and $T$ is the positive delay time.
The evolution of the dynamics is dependent on not only the present value $x(t)$ but also earlier one $x(t-T)$.
To solve equation~\eqref{eq:delay}, it is discretized with the following way:
1) each interval $T$ is divided into $N$ subintervals $h$ and each subinterval $h$ is approximated with a scalar value, where $T=N\times h$;
2) the $N$ samples of each interval are considered as an $N$-dimension vector.
In \cite{Mannai:delay:ND15}, an image encryption algorithm based on a time-delay Ikeda system (IEATD) was proposed.
The designers of IEATD believed that utilizing the rich dynamics of a discretized Ikeda system and a new keystream generation mechanism associated with
the average of all pixels of the plain-image can provide sufficient capacity to withstand known/chosen-plaintext attacks.

In reality, the security strength of IEATD is very weak as its equivalent secret key can be obtained with only two chosen plain-images \cite{lim:improve:IM18}.
Meanwhile, M. Li et al. pointed out that two security defects exist in IEATD:
1) the regularity of the keystream and absence of position scrambling;
2) incapacity to resist differential attack \cite{lim:improve:IM18}.
To remedy the defects, they adopted much more complex encryption operations: permutation and crossover diffusion phases.
In short, we call the enhanced image encryption algorithm using the crossover diffusion as IEACD.
This paper focuses on security analysis of the two image encryption algorithms, IEATD \cite{Mannai:delay:ND15} and IEACD \cite{lim:improve:IM18}.
We found that the authors of \cite{lim:improve:IM18} did not notice a fatal drawback of the keystream generation mechanism: insensibility to minor change of a pixel.
This leads to that IEACD still cannot withstand chosen-plaintext attack.
Furthermore, there is improper keystream configuration in diffusion that almost discloses the whole keystream.
The essential structures of the two algorithms cause that the equivalent secret key of IEATD and IEACD can be recovered with
known-plaintext attack and chosen-plaintext attack, respectively.

The rest of this paper is organized as follows.
Section~\ref{sec:algorithm} concisely describes the encryption procedures of the encryption algorithm IEATD and its enhanced version
IEACD. Then, Sec.~\ref{sec:analysiIEATD} and Sec.~\ref{sec:CryptofIEACD} present the detailed cryptanalysis results on the two encryption algorithms, respectively.
The last section concludes the paper.

\section{Description of two analyzed image encryption algorithms}
\label{sec:algorithm}

The input of algorithm IEATD is an 8-bit gray-scale image of size $H \times W$. The plain-image is scanned in the raster order and then can be represented as a sequence $\mathbf{I}=\{I(i)\}_{i=0}^{HW-1}$.
The corresponding cipher-image is denoted by $\mathbf{I}'=\{I'(i)\}_{i=0}^{HW-1}$.
Then, IEATD and its enhanced version IEACD can be described in Sec.~\ref{ssec:IEATD} and Sec.~\ref{ssec:IEACD}, respectively.

\subsection{The framework of IEATD}
\label{ssec:IEATD}

\begin{itemize}[leftmargin=*]
\item \textit{The secret key}: a positive integer $N_0$, three control parameters of discretized Ikeda chaotic system
      \begin{equation}
      \label{eq:Ikeda}
      	\left\{
      \begin{aligned}
        X_{1}(k+1) &= X_{2}(k) \\
        X_{2}(k+1) &= X_{3}(k) \\
                   & \vdots \\
        X_M(k+1)   &= X_M(k) + h \cdot(-\alpha\cdot X_M(k) + m\cdot \sin(X_1(k)))
      \end{aligned}
      \right.
      \end{equation}
      and its initial condition $\mathbf{X}(0) = \{X_{i}(0)\}_{i=0}^{M-1}$, where $h\in[0.1, 1]$, $\alpha\in [1, 6]$, and $m \in [18, 20]$.

  \item \textit{The confusion procedure}:

  \begin{itemize}[leftmargin=*]
    \item \textit{Step 1}: Divide $\{I(i)\}_{i=0}^{HW-1}$ into $s$ vectors, where the length of $k$-th vector $I^{(k)}$ is $N(k)$, and
    the length of subsequent vector depends on the previous one:
    \begin{equation} \label{eq:length}
    N(k) = N(k-1) + \lfloor \text{mean}(I^{(k-1)}) \rfloor,
    \end{equation}
    where $k=1\sim (s-2)$, $N(0)=N_0$. Finally, assign the actual length of the last vector $I^{(s-1)}$ to $N(s-1)$.
    Obviously, the longest vector is either the last vector $I^{(s-1)}$ or the penultimate one $I^{(s-2)}$.

    \item \textit{Step 2}: As for the $k$-th vector $I^{(k)}$, iterate Eq.~\eqref{eq:Ikeda} $N(k)$ times from the initial condition and generate a chaotic sequence  $S^{(k)}=\{S^{(k)}(i)\}_{i=0}^{N(k)-1}$,
        and then obtain the quantized sequence $Y^{(k)}=\{Y^{(k)}(i)\}_{i=0}^{N(k)-1}$ via
        \begin{equation*}
        Y^{(k)}=\left(S^{(k)}\cdot 10^{14} \right) \bmod 256,
        \end{equation*}
        where $k=0\sim (s-1)$.
        Finally, concatenate $\{Y^{(k)}\}_{k=0}^{s-1}$ further into a sequence $\mathbf{Y} = \{Y(i)\}_{i=0}^{HW-1}$, where
        \[
        Y\left( i+\sum_{j=0}^{k-1}N(j) \right)=Y^{(k)}(i).
        \]
    The longest vector among $\{Y^{(k)}\}_{k=0}^{s-1}$ is either $Y^{(s-1)}$ or $Y^{(s-2)}$, and let $Y^{\rm L}$ denote it.
    Note that every element of $\{Y^{(k)}\}_{k=0}^{s-1}$ is a subsequence of $Y^{\rm L}$.

    \item \textit{Step 3}: Perform confusion operations on sequence $\mathbf{I}$ by
        \begin{equation}
        \label{eq:confu}
            I'(i)=I(i) \oplus Y(i)
        \end{equation}
        for $i=0 \sim (HW-1)$, where $\oplus$ donates the bitwise XOR operation.
  \end{itemize}
\end{itemize}

\subsection{The enhanced elements of IEACD compared with IEATD}
\label{ssec:IEACD}

To enhance the security level of IEATD, some extra operations were appended to withstand the chosen-plaintext attack reported in \cite{lim:improve:IM18}.

\begin{itemize}[leftmargin=*]
\item \textit{The added secret sub-keys}: a positive integer $C$, the initial condition $q(0)\in (0, 1)$ and the control parameter $\beta \in (3.5699456,4]$ of Logistic map
      \begin{equation}
      \label{eq:Logistic}
      q(i+1)=\beta \cdot q(i) \cdot (1-q(i)).
      \end{equation}

  \item \textit{The modified encryption procedures}:

  \textit{Step 1}: Iterate Eq.~\eqref{eq:Logistic} $HW$ steps from $q(0)$, and obtain a chaotic sequence $\mathbf{Q}=\{q(i)\}_{i=0}^{HW-1}$, which
  is used to produce permutation vector $\mathbf{Z}=\{z(i)\}_{i=0}^{HW-1}$, where $q(i)$ is the $z(i)$-th largest element in the sequence $\mathbf{Q}$.
  Then, permute $\mathbf{I}$ with the permutation vector $\mathbf{Z}$ and obtain a permuted intermediate image $\mathbf{I}^{\star}=\{I^{\star}(i)\}_{i=0}^{HW-1}$ by performing
  \begin{equation} \label{eq:permute1}
  I^{\star}(i) = I(z(i))
  \end{equation}
  for $i=0 \sim (HW-1)$.

  \textit{Step 2}: Divide $\{I^{\star}(i)\}_{i=0}^{HW-1}$ into $s$ vectors $\{I^{\star(k)}\}_{k=0}^{s-1}$ and obtain a sequence $\mathbf{Y}$ like \textit{Step 1, 2} of IEATD, where the division size is determined by
  \begin{equation}\label{eq:length2}
    N(k) = N(k-1) + \lfloor \text{mean}(I^{\star(k-1)}) \rfloor.
  \end{equation}
  Then perform the confusion operation on $\{I^{\star}(i)\}_{i=0}^{HW-1}$ via
  \begin{equation} \label{eq:confusion}
    I^{\star\star}(i) = I^{\star}(i) \oplus Y(i),
  \end{equation}
  where $i=0, 1, \cdots, (HW-1)$.

  \textit{Step 3}: Crossover diffusion:
  \begin{itemize}
    \item \textit{Step 3a}: Generate index array $\mathbf{U}=\{u(i)\}_{i=0}^{HW-1}$, where
    \begin{equation}\label{eq:cal-ind}
        u(i)=
        \begin{cases}
            \lfloor \frac{i}{2} \rfloor                                & \text{if } i \bmod 2 = 0; \\
            \lfloor \frac{i}{2} \rfloor + \lfloor \frac{HW}{2} \rfloor & \text{if } i \bmod 2 = 1,
        \end{cases}
    \end{equation}
    $i=0, 1, \cdots, (HW-1)$.

    \item \textit{Step 3b}: Conduct the first round crossover diffusion with $C$ and sequence $\mathbf{Z}$ by
    \begin{equation}
    \label{eq:cdiff1}
        I^{*}(u(i))=I^{\star\star}(u(i)) \oplus \left( I^{*}(u(i-1)) \boxplus z(u(i)) \right)
    \end{equation}
    for $i=1 \sim (HW-1)$, where $a \boxplus b=(a+b) \bmod 256$
    and $I^{*}(u(0))=I^{\star\star}(u(0)) \oplus \left(C \boxplus z(u(0))\right)$.

    \item \textit{Step 3c}: Perform the second round of crossover diffusion via
    \begin{equation}
    \label{eq:cdiff2}
        I^{**}(u(i)) = I^{*}(u(i)) \oplus \left( I^{**}(u(i-1)) \boxplus z(u(i)) \right)
    \end{equation}
    for $i=1 \sim (HW-1)$, where $I^{**}(u(0))=I^{*}(u(0))\oplus \left( I^{*}(u(HW-1)) \boxplus z(u(0)) \right)$.
  \end{itemize}

  \textit{Step 4}: Permute $\{I^{**}(i)\}_{i=0}^{HW-1}$ and obtain the ciphertext $\{I'(i)\}_{i=0}^{HW-1}$ by performing
  \begin{equation}
  \label{eq:permute2}
  I'(z(i))=I^{**}(i)
  \end{equation}
  for $i=0\sim (HW-1)$.
\end{itemize}

\section{Cryptanalysis of IEATD}
\label{sec:analysiIEATD}

In \cite{Mannai:delay:ND15}, the authors claimed that the adopted intermediate keystream is dependent on the plaintext, so it can withstand the classic attacks, such as plain/chosen-plaintext attack and chosen-ciphertext attack. However, we argue that the statement is not always correct.
In this section, the weak keys about $N_0$ are discussed.
After briefly describing the chosen-plaintext attack on IEATD proposed in \cite{lim:improve:IM18}, we present a known-plaintext attack on it.

\subsection{Weak keys with respect to $N_0$}
\label{sec:IEATD:weak-key}

\begin{figure}[!htb]
  \centering
  \begin{minipage}{\ThreeImW}
    \centering
    \includegraphics[width=\ThreeImW]{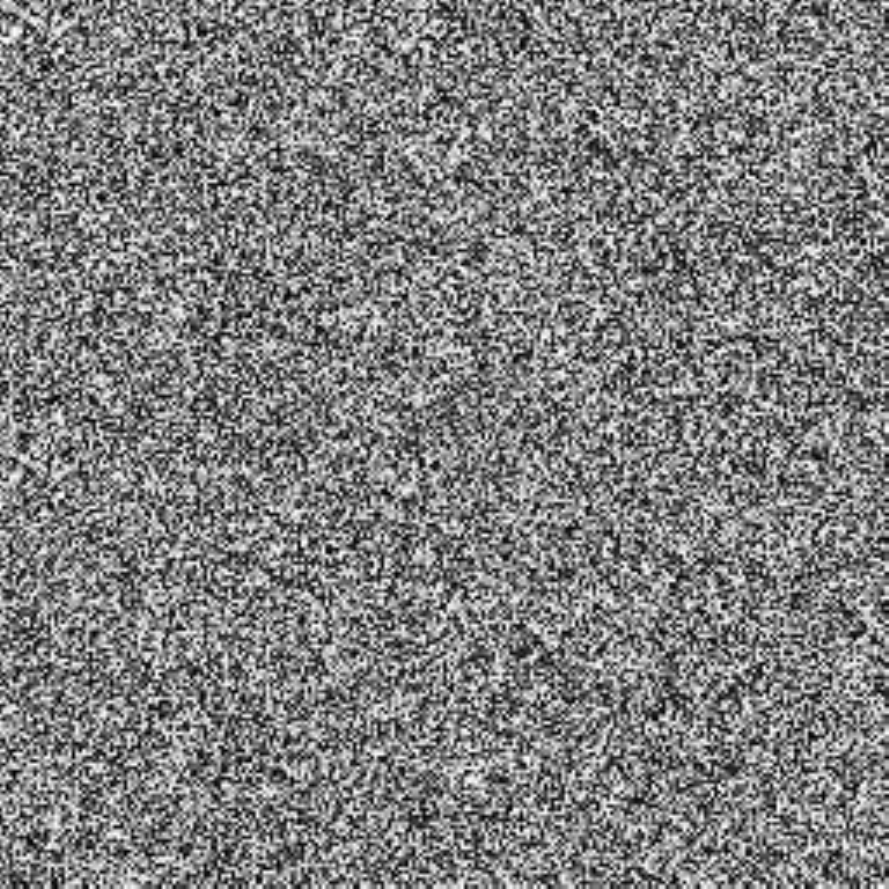}
    a)
  \end{minipage}
  \begin{minipage}{\ThreeImW}
    \centering
    \includegraphics[width=\ThreeImW]{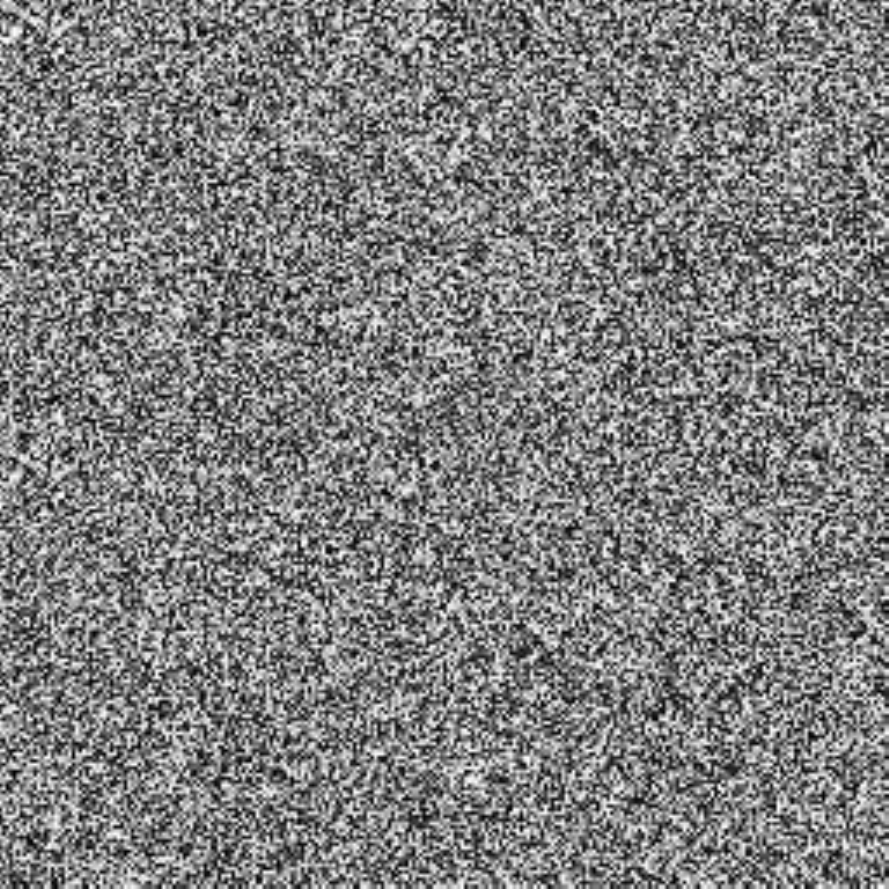}
    b)
  \end{minipage}
  \begin{minipage}{\ThreeImW}
    \centering
    \includegraphics[width=\ThreeImW]{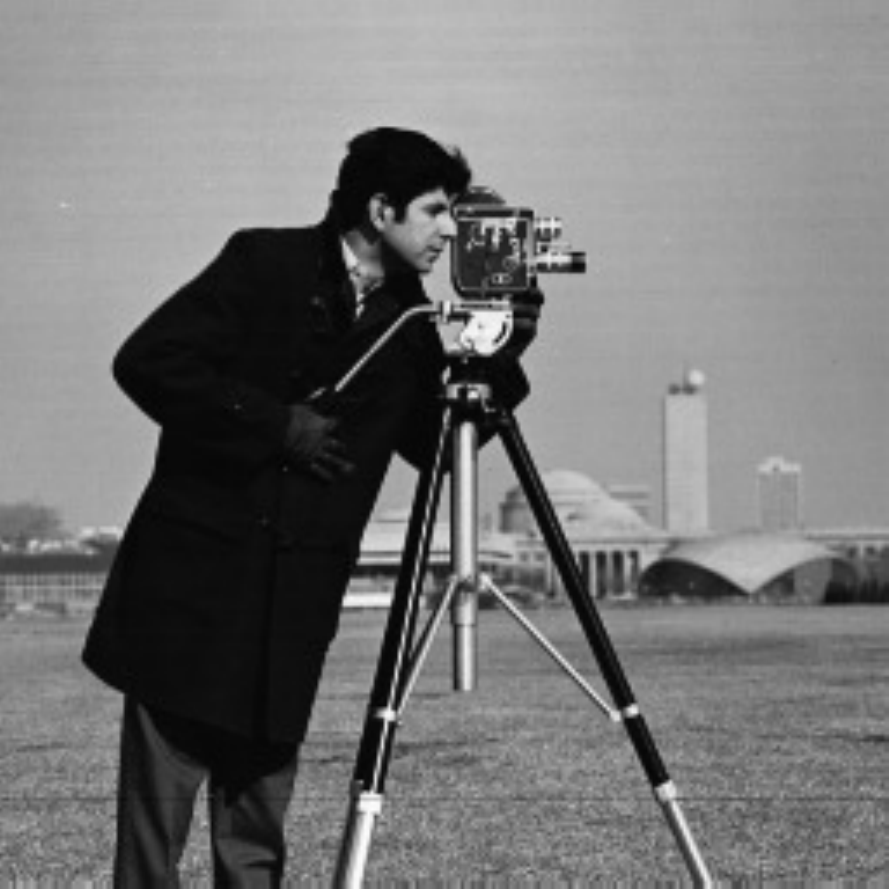}
    c)
  \end{minipage}
  \caption{The decrypted result using a mask image when $N_0 \geq L$: a) the mask image; b) a cipher-image; c) the decrypted result.}
\label{fig:weak}
\end{figure}

In IEATD, the plaintext is first divided into some vectors, which is dependent on a given key $N_0$ and the plaintext itself.
The scope of $N_0$ is not specifically given in the algorithm. However, $N_0$ should be less than $L$ from the security standpoint, where $L=\lfloor \frac{HW}{2} \rfloor$.
If $N_0\geq L$, it is observed that the keystream generation mechanics is futile since sequence $\mathbf{Y}$ generated from any plain-image is the same.
In such case, a mask image, generated by XORing a plain-image and the corresponding cipher-image pixel by pixel, can be directly used as the equivalent secret key.
As shown in Fig.~\ref{fig:weak}, a cipher-image is fully decrypted with the mask image.
As this is contrary to the original intention of the designers, it is assumed that $N_0<L$ in the subsequent analysis.

\subsection{The chosen-plaintext attack proposed by M. Li et al.}
\label{sec:IEATD:cpa}

To make the cryptanalysis of IEATD more complete, we briefly introduce the chosen-plaintext attack on IEATD proposed by M. Li et al. in \cite{lim:improve:IM18} and
comment its performance:
\begin{itemize}
  \item Determining $N_0$: Referring to Eq.~\eqref{eq:length}, one can see that $N(i)=N_0$ and $Y^{(i)}=Y^{(0)}$ for any $i\in \{1, 2, \cdots, s-2\}$ if $\lfloor\text{mean}(I^{(k)})\rfloor =0$ for $k=0 \sim (s-2)$.
      To ensure this condition exists, one can choose a plain-image of fixed value zero. Then, sequence $\mathbf{Y}$ can be obtained by $\mathbf{Y}=\mathbf{I}\oplus \mathbf{I}'$.
      If one calculates the autocorrelation coefficients $R(t)$ of $\mathbf{Y}$, the maximum should be $R(N_0)$, where
      \begin{equation*}
        R(t) = \frac{ \sum_{i=0}^{HW-t-1} (Y(i)-\bar{\mu})\cdot (Y(i+t)-\bar{\mu}) }{ \sum_{i=0}^{HW-1} (Y(i)-\bar{\mu})^2 },
      \end{equation*}
      $\bar{\mu}$ is the average of $\mathbf{Y}$, and $t\in [1, L-1]$ is lag.

  \item Obtaining $Y^{\rm L}_{255}$: Choose a plain-image $\mathbf{I}_{255}$ of fixed value 255 and get the corresponding longest sequence $Y^{\rm L}_{255}$ from $\mathbf{Y}_{255}$,
  where $\mathbf{Y}_{255}$ is the result by XORing $\mathbf{I}_{255}$ and its cipher-image $\mathbf{I}_{255}'$ pixel by pixel.
\end{itemize}

Set $Y^{\rm L}=Y^{\rm L}_{255}$ and denote the length of $Y^{\rm L}$ by $N_{Y^{\rm L}}$. The decryption procedure can be described as follows:
\begin{itemize}
  \item \textit{Step 1}: Set $N(0)=N_0$, $N_{\rm sum}=0$, $k=0$.

  \item \textit{Step 2}: Do $I(N_{\rm sum}+i)=I'(N_{\rm sum}+i)\oplus Y^{\rm L}(i)$ for $i=0\sim (N(k)-1)$, and then set $k=k+1$.

  \item \textit{Step 3}: Set
  \[
  N_{\rm sum}=N_{\rm sum}+N(k-1)
  \]
   and calculate $N(k)$ via Eq.~\eqref{eq:length}.
    If
    \begin{equation}\label{eq:condition}
    N(k) \leq N_{Y^{\rm L}}
    \end{equation}
    and $N_{\rm sum}\leq HW-N(k)$, go back to \textit{Step 2}.

  \item \textit{Step 4}: Assign $\min((HW-N_{\rm sum}), N_{Y^{\rm L}})$ to $N(k)$, where $\min(a, b)$ returns the smaller element between
 $a$ and $b$. Then do
 \[
 I(N_{\rm sum}+i)=I'(N_{\rm sum}+i)\oplus Y^{\rm L}(i)
 \]
 for $i=0\sim (N(k)-1)$.
\end{itemize}

In the above decryption process, the intermediate keystream $\mathbf{Y}=\{Y(i)\}_{i=0}^{HW-1}$ corresponding to a cipher-image is gradually recovered.
In other words, one can reconstruct the specific $\mathbf{Y}$ belonging to a cipher-image from $Y^{\rm L}_{255}$ and $N_0$.
Thus, they can be regarded as the equivalent secret key.
The time complexity of the attack is $O((WH)^2)$ instead of that claimed in \cite{lim:improve:IM18}, $O(WH)$.

\subsection{Known-plaintext attack on IEATD}
\label{sec:IEATD:kpa}

Known-plaintext attack can be considered as a stronger version of the chosen-plaintext attack as the former can recover the information with the secret key from some given plaintexts, instead of that specially constructed or selected.
As for algorithm IEATD, even if only one plain-image and the corresponding cipher-image are available, one can obtain $N_0$ effortlessly and then derive a counterpart of $Y^{\rm L}$, $Y^{\rm L'}$.
They can be used to disclose some visual information of the other cipher-images encrypted with the same secret key.

According to Eq.~\eqref{eq:length}, one can get $\{Y^{(k)}\}_{k=0}^{s-1}$ after obtaining $N_0$ and $\mathbf{Y}$.
Then, as for two adjacent vectors $Y^{(k)}$ and $Y^{(k+1)}$, one has $Y^{(k)}(i)=Y^{(k+1)}(i)$ for $i=0\sim (N(k)-1)$, where $k=0\sim (s-2)$. Hence, the condition can be used to verify
the search of $N_0$.
Since the scope of $N_0$ is relatively small as mentioned in Sec.~\ref{sec:IEATD:weak-key},
the confirmation of $N_0$ is feasible through brute-force searching:
\begin{itemize}
  \item \textit{Step 1}: Produce sequence $\mathbf{Y}$ by XORing the plain-image and its corresponding cipher-image pixel by pixel.

  \item \textit{Step 2}: For $j=0 \sim (L-1)$, do the following operations:
  \begin{itemize}
    \item \textit{Step 2a}: If condition $\{Y(i)\}_{i=0}^{j-1}=\{Y(i)\}_{i=j}^{2j-1}$
        satisfies, set $N(0)=j$, $N_{\rm sum}=0$, $k=1$; otherwise, go to the next loop.

    \item \textit{Step 2b}: Set $N_{\rm sum}=N_{\rm sum}+N(k-1)$ and calculate $N(k)$ using Eq.~\eqref{eq:length}.
        If $N_{\rm sum}>HW-2N(k)$, set $N_0=j$ and terminate the attack.

    \item \textit{Step 2c}: If condition
        \begin{equation*}
          \left\{ Y(i) \right\}_{i=N_{\rm sum}}^{N_{\rm sum}+N(k)-1}=\{Y(i)\}_{i=N_{\rm sum}+N(k)}^{N_{\rm sum}+2N(k)-1}
        \end{equation*}
        satisfies, set $k=k+1$ and go back to \textit{Step 2b}.
  \end{itemize}
\end{itemize}

\begin{figure}[!htb]
  \centering
  \begin{minipage}{\ThreeImW}
    \centering
    \includegraphics[width=\ThreeImW]{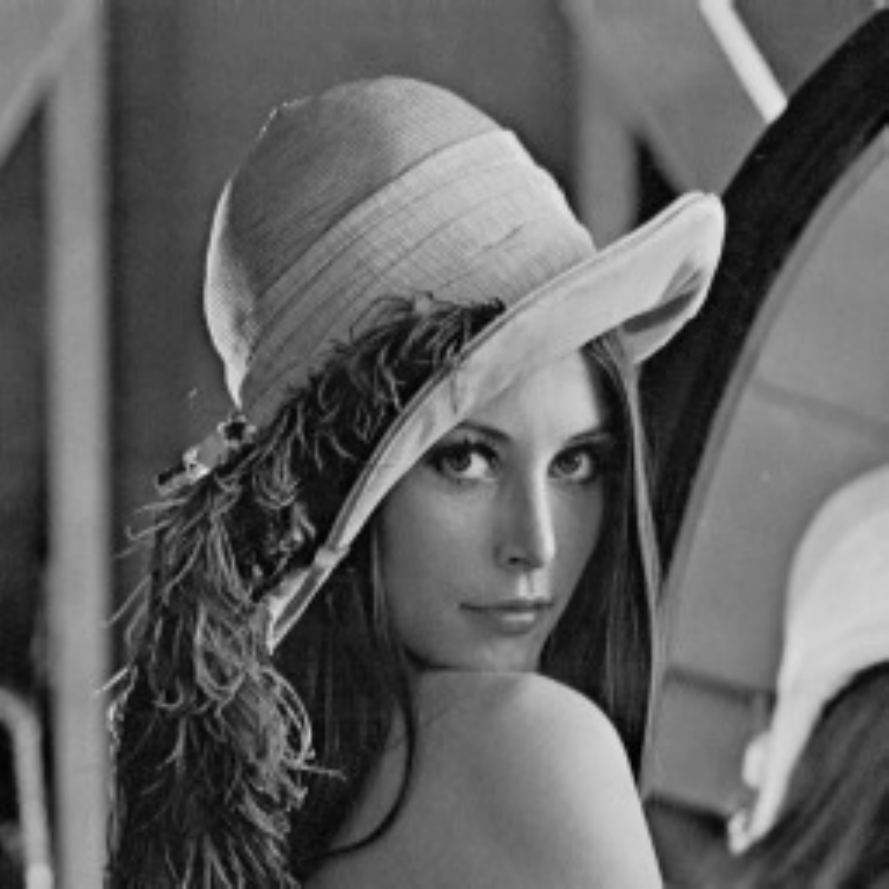}
    a)
  \end{minipage}\hspace{\figsep}
  \begin{minipage}{\ThreeImW}
    \centering
    \includegraphics[width=\ThreeImW]{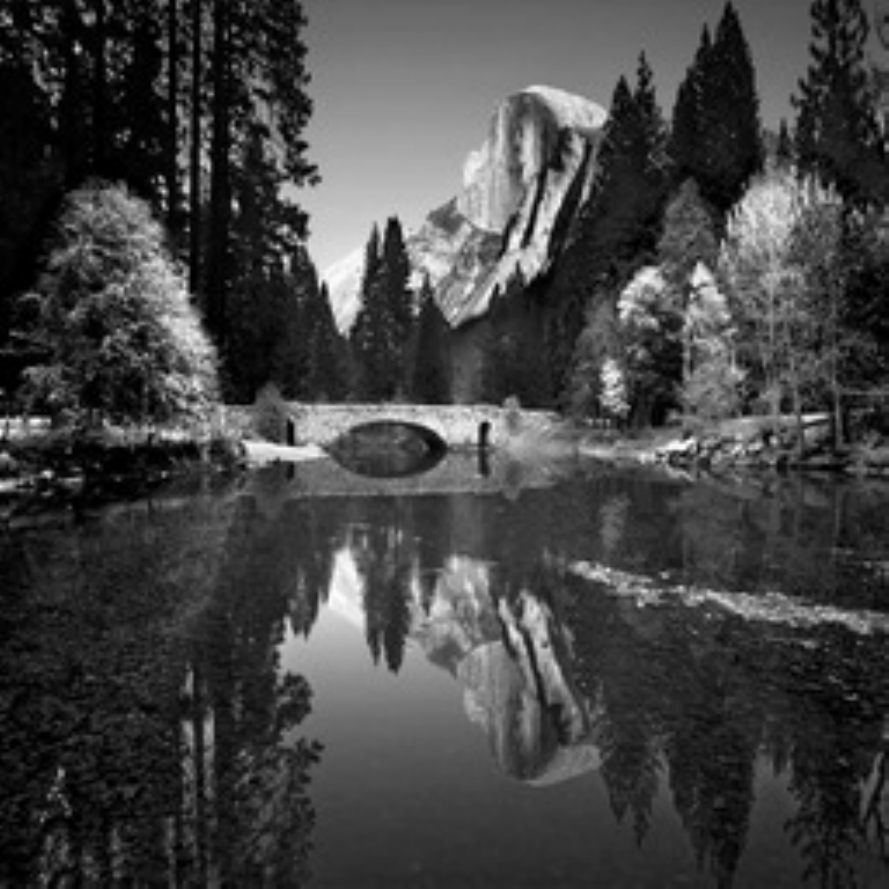}
    b)
  \end{minipage}\hspace{\figsep}
  \begin{minipage}{\ThreeImW}
    \centering
    \includegraphics[width=\ThreeImW]{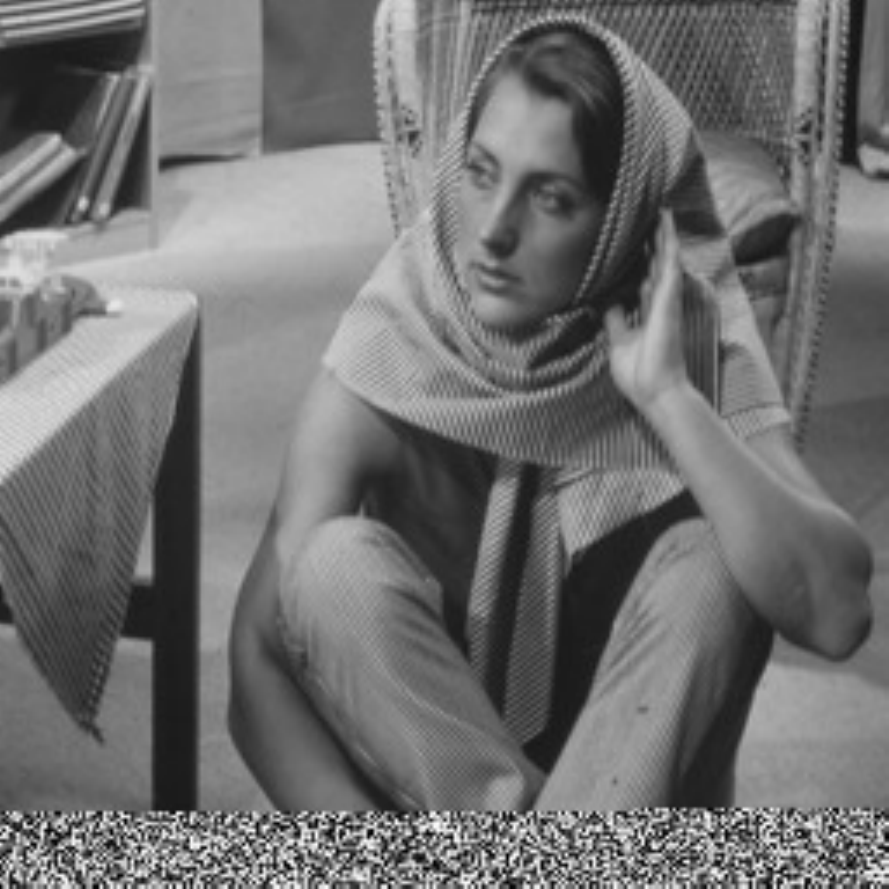}
    c)
  \end{minipage}\hspace{\figsep}
  \begin{minipage}{\ThreeImW}
    \centering
    \includegraphics[width=\ThreeImW]{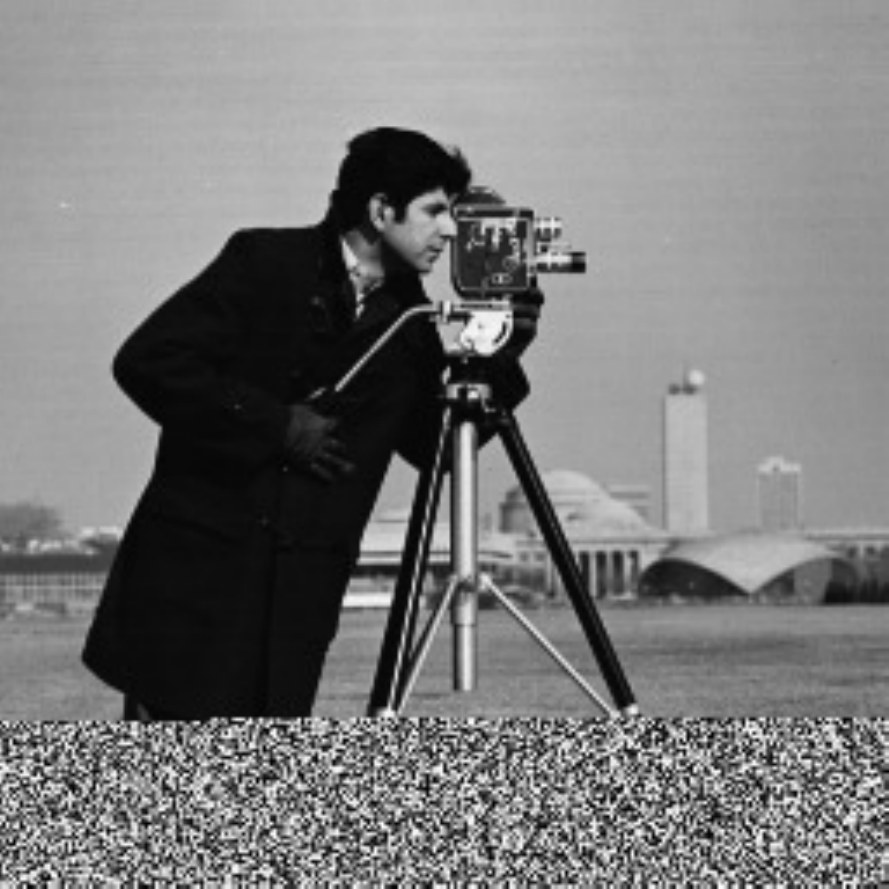}
    d)
  \end{minipage}\\
  \begin{minipage}{\ThreeImW}
    \centering
    \includegraphics[width=\ThreeImW]{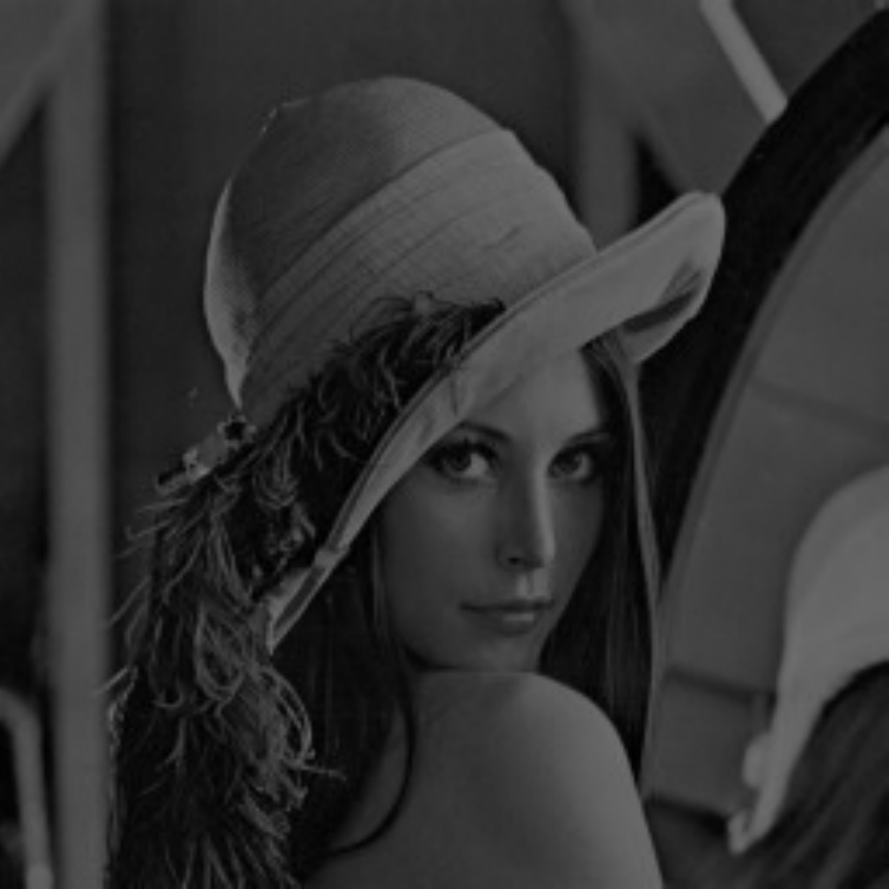}
    e)
  \end{minipage}\hspace{\figsep}
  \begin{minipage}{\ThreeImW}
    \centering
    \includegraphics[width=\ThreeImW]{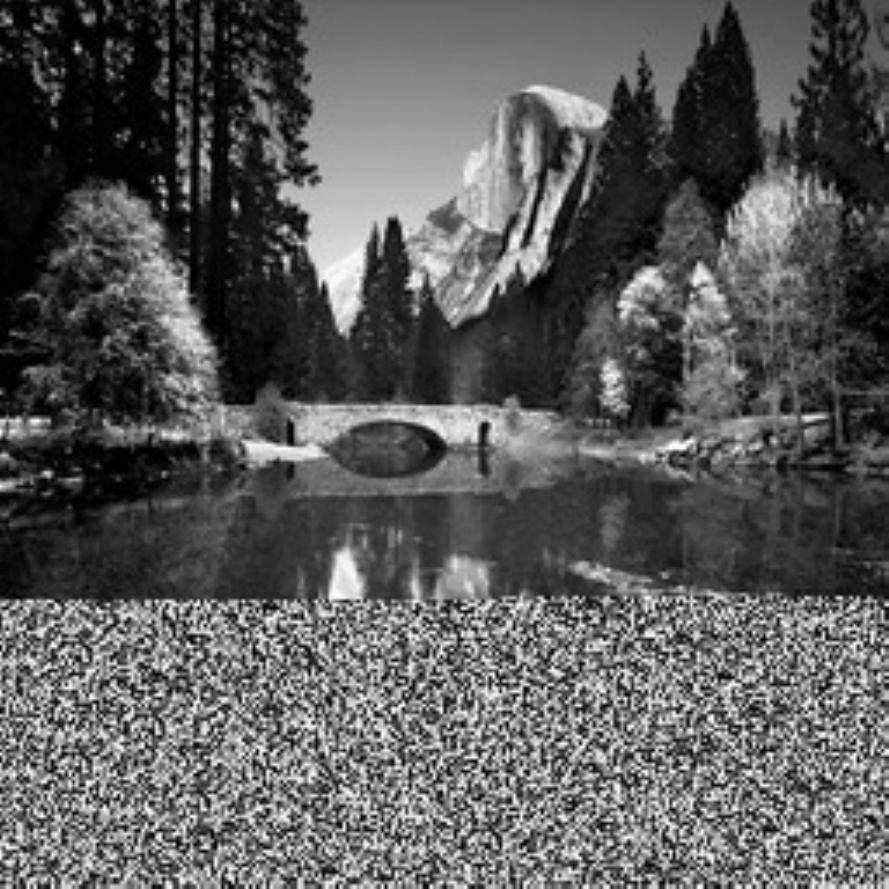}
    f)
  \end{minipage}\hspace{\figsep}
  \begin{minipage}{\ThreeImW}
    \centering
    \includegraphics[width=\ThreeImW]{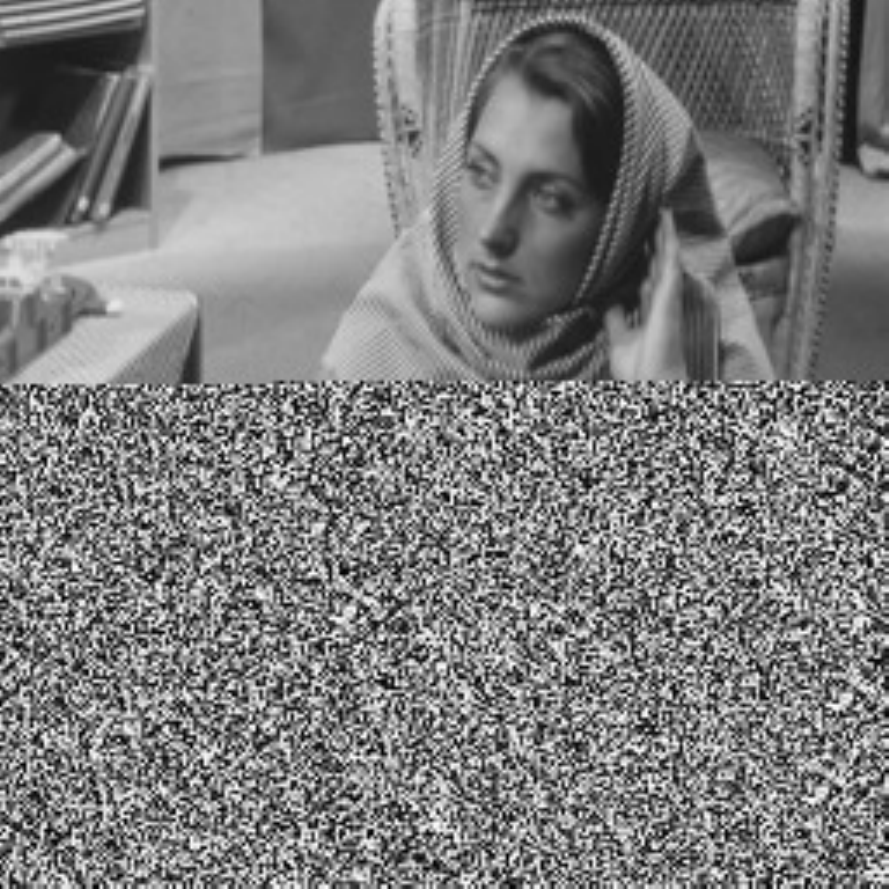}
    g)
  \end{minipage}\hspace{\figsep}
  \begin{minipage}{\ThreeImW}
    \centering
    \includegraphics[width=\ThreeImW]{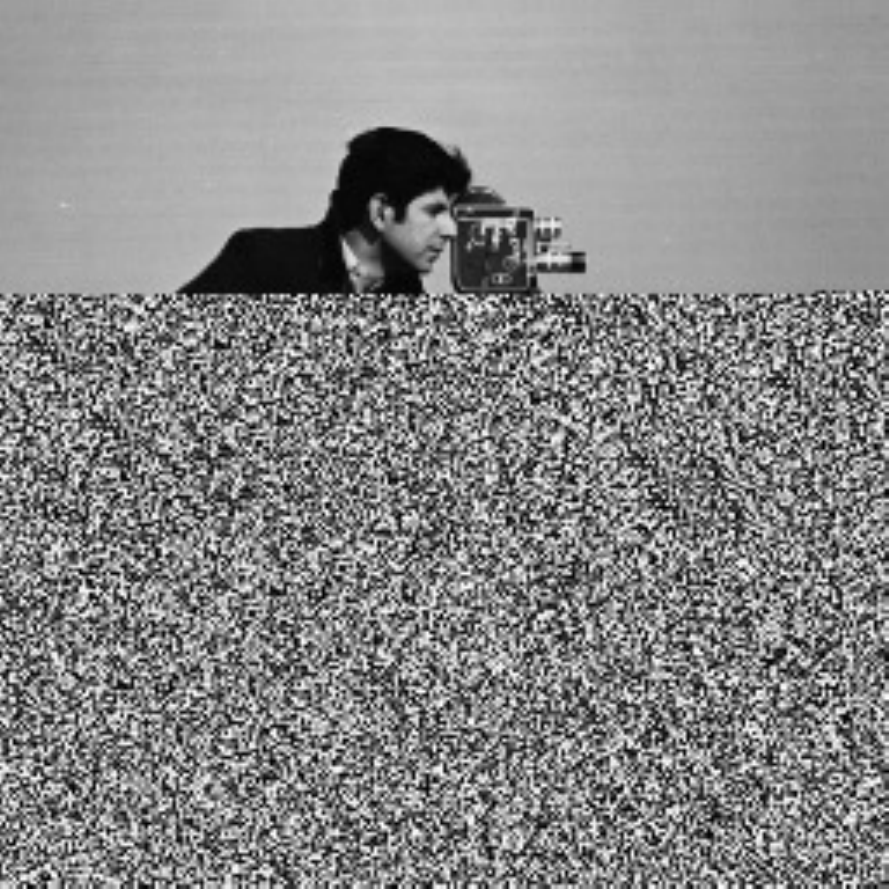}
    h)
  \end{minipage}
\caption{The result of known-plaintext attack on IEATD:
a) a known plain-image; b)-d) the decrypted images with the equivalent secret key derived from Fig.~a);
e) a known plain-image with half brightness of that in Fig.~\ref{fig:reuse}a); f)-h) the decrypted images with the key obtained from Fig.~\ref{fig:reuse}e).}
\label{fig:reuse}
\end{figure}

After confirming $N_0$, one can easily obtain $Y^{\rm L'}$ from $\mathbf{Y}$.
Set $Y^{\rm L}=Y^{\rm L'}$ and the decryption procedure is the same as that in Sec.~\ref{sec:IEATD:cpa}.
Once condition~\eqref{eq:condition} does not exist during the decryption process, the following cipher-pixels are all decrypted incorrectly.
Referring to Eq.~\eqref{eq:length}, one can know a simple rule:
the brighter a plain-image is, the fewer the divided vectors become (the corresponding $Y^{\rm L}$ gets longer).
When the brightness of the plain-image corresponding to a cipher-image to be decrypted is higher than that of the known plain-image,
condition~\eqref{eq:condition} is not satisfied for smaller index $k$.
This means that more portion of the cipher-image cannot be decrypted correctly.
As shown in Fig.~\ref{fig:reuse}b), the image with lower brightness than that in Fig.~\ref{fig:reuse}a) is even completely decrypted.
By contrast, many consecutive pixels of two brighter plain-images cannot be recovered correctly (See Fig.~\ref{fig:reuse}c) and d)).
As for the same plain-image to be decrypted, if brightness of the available known plain-image is lower, more consecutive pixels cannot be decrypted correctly.
This point can be verified by comparing Fig.~\ref{fig:reuse}b), c) and d) with Fig.~\ref{fig:reuse}f), g) and h), respectively.

\section{Cryptanalysis of IEACD}
\label{sec:CryptofIEACD}

To cope with the insecurity problems of IEATD reported in \cite{lim:improve:IM18}, multiple confusion and diffusion operations are appended, making the algorithm become another algorithm IEACD.
In fact, the original keystream generation mechanism indeed exists a serious pitfall, which leads to that the patched algorithm IEACD still cannot withstand chosen-plaintext attack.
In this section, three weaknesses of IEACD are first analyzed to facilitate description of the following chosen-plaintext attack.

\subsection{Three weaknesses of IEACD}
\label{sec:IEACD:weak}

\begin{itemize}

\item The real size of key space is much smaller than the expected one

Due to the limitation of finite-precision presentation, dynamics of any chaotic system is definitely degraded when it is
implemented in a digital device. As investigated in \cite{Ding:period:ND19,cqli:network:TCASI2019,cqli:Cat:TC21}, the structure of
the state-mapping network (SMN) of a digitized chaotic system implemented with fixed-point precision $e+1$ is largely dominated by
that with precision $e$. The short period problems of PRNG based on Logistic map (\ref{eq:Logistic}) implemented in a digital device (with
fixed-point arithmetic or floating-point arithmetic) were comprehensively discussed in \cite{cqli:network:TCASI2019}.
As shown in Fig.~\ref{fig:SMNs}, discretized Ikeda system obeys this rule also.
No matter what the precision is, the SMN of discretized Ikeda system follows the following rules:
1) an SMN is composed of some weakly connected components;
2) there are some self-loops (an edge connecting a node to itself);
3) As for each connected component, there is one and only one cycle (including special cycle, self-loop), and every node evolves to it via a transient process;
4) Many nodes have two and only two parent nodes.
Generating a pseudo-random number sequence by the orbits determined by a chaotic map is actually walking along a path of an SMN.
Now, one can see that the period of a sequence by solving the discretized Ikeda system may be very short (even only one).
So, there are a number of equivalent secret keys and invalid secret keys as for the function of IEATD and IEACD.
Note that such pitfall always exists no matter how large the precision $e$ gets.

\begin{figure}[!htb]
  \centering
  \begin{minipage}{\BigOneImW}
    \centering
    \includegraphics[width=\BigOneImW]{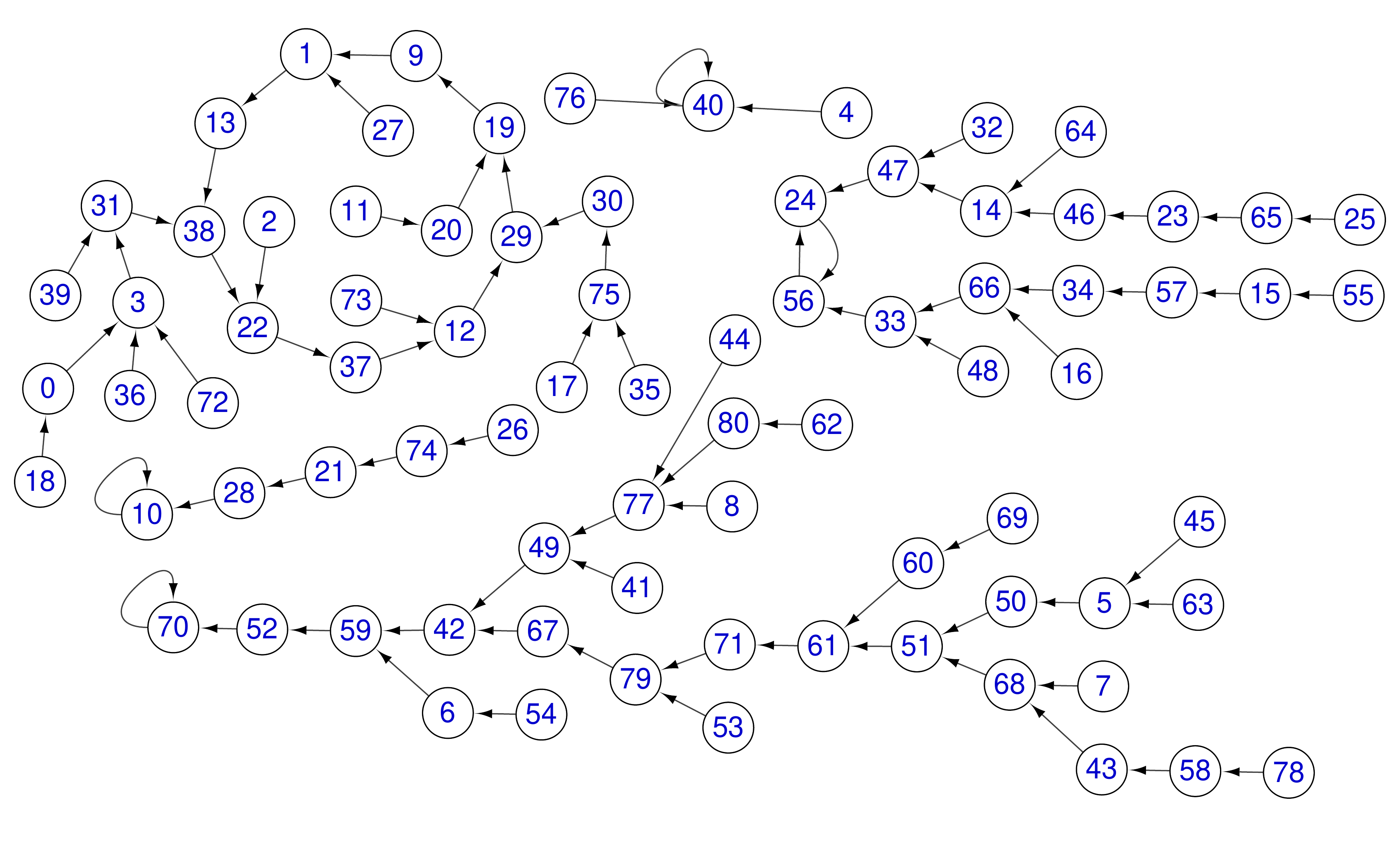}
    a)
  \end{minipage}
  \begin{minipage}{1.5\BigOneImW}
    \centering
    \includegraphics[width=1.6\BigOneImW]{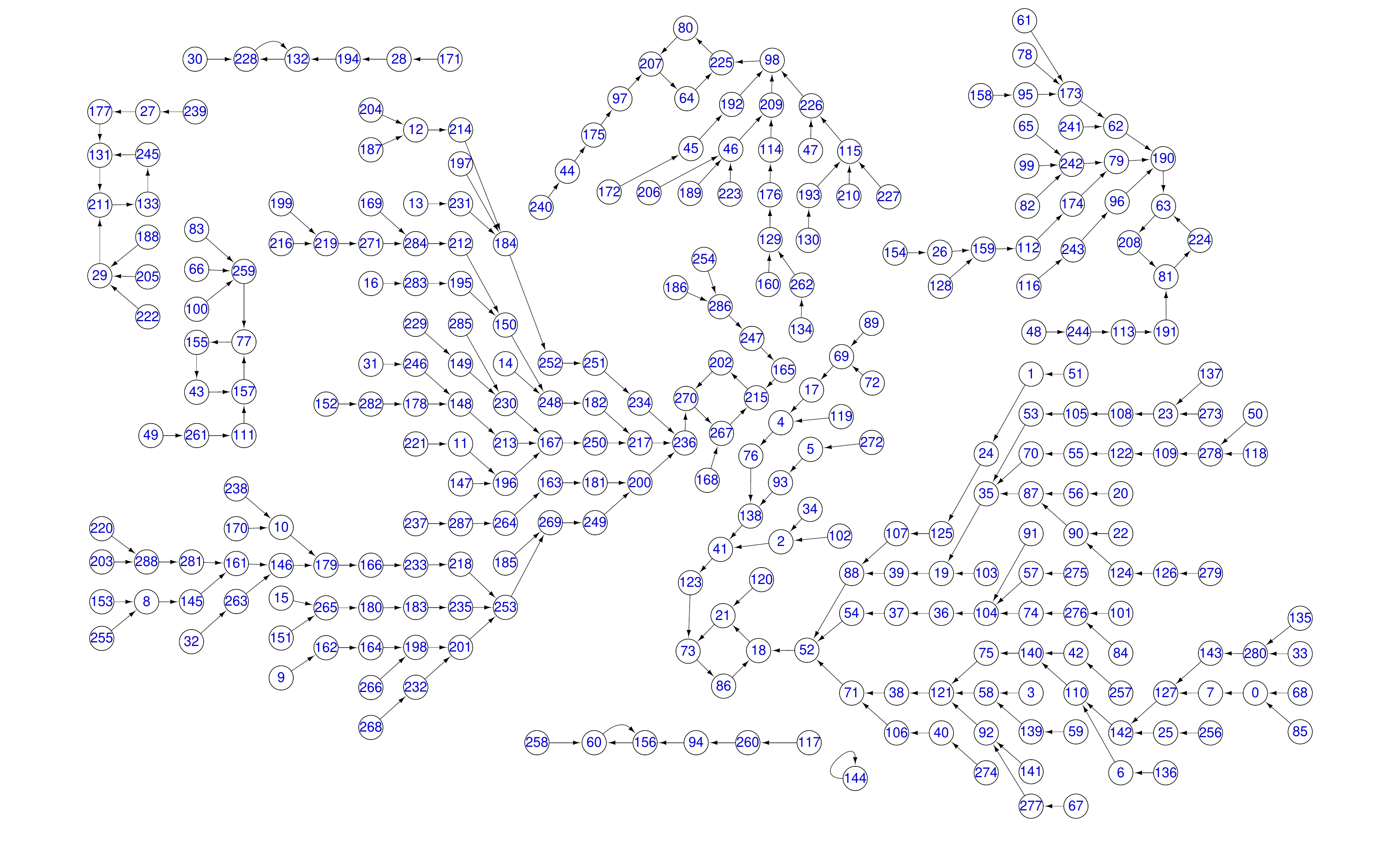}
    b)
  \end{minipage}
  \caption{The State-Mapping Networks of the Ikeda map with $\alpha=\frac{48}{2^3}, h=\frac{1}{2^3}, m=\frac{156}{2^3}$ under $e$-bit fixed-point precision: a) $e=3$; b) $e=4$.}
  \label{fig:SMNs}
\end{figure}

\item Insensibility of keystream generation mechanics

Although the permutation operations are performed before the confusion step to frustrate the predictability of keystream $\mathbf{Y}$, the keystream is still insensitive to minor changes of some pixels
of the plain-image. Referring to Eq.~\eqref{eq:length2}, one can see that the possibility that $N(k+1)$ change is roughly $\frac{1}{N(k)}$ when a pixel in vector $I^{\star(k)}$ is slightly changed with variation $\Delta=1$.
More generally, when a randomly chosen pixel in one plain-image is slightly altered with variation $\Delta=1$, the probability that $\mathbf{Y}$ generated by the altered plain-image is different from the previous version is
\begin{equation*}
P_{\rm c}=\sum_{k=0}^{s-1}\left(\frac{1}{N(k)} \cdot \frac{N(k)}{HW} \right)=\frac{s}{HW}.
\end{equation*}
Taking a nature image of size $256\times 256$ as an example, $s\approx 40$ and $P_{\rm c} \approx 6.103 \times 10^{-4}$.
Note that the more the average of the pixels of the image approaches 255, the smaller the probability is.

\item Improper configuration of keystream

The two-round crossover diffusion is performed to resist chosen-plaintext attack and differential attack, but the keystream $\mathbf{Z}$ used in permutation is wrongly reused in the diffusion part, which makes the algorithm more insecure.
From Eq.~\eqref{eq:permute1}, \eqref{eq:confusion} and \eqref{eq:cdiff1}, one has
\begin{align} \label{eq:first}
  I^{*}(u(i)) &= I^{\star\star}(u(i)) \oplus \left(I^{*}(u(i-1)) \boxplus z(u(i))\right) \nonumber\\
           &= I^{\star}(u(i)) \oplus Y(u(i)) \oplus \left(I^{*}(u(i-1)) \boxplus z(u(i))\right) \nonumber\\
           &= I(z(u(i))) \oplus Y(u(i)) \oplus \left(I^{*}(u(i-1)) \boxplus z(u(i))\right),
\end{align}
where $i>0$ and $I^{*}(u(0))=I(z(u(0))\oplus Y(u(0))\oplus \left(C \boxplus z(u(0))\right)$.
Also, incorporating $i=u(i)$ and Eq.~\eqref{eq:permute2} into Eq.~\eqref{eq:cdiff2}, one has
\begin{align}
\label{eq:second}
  I'(z(u(i))) &= I^{**}(u(i)) \nonumber\\
              &= I^{*}(u(i)) \oplus \left(I^{**}(u(i-1)) \boxplus z(u(i))\right) \nonumber\\
              &= I^{*}(u(i)) \oplus \left(I'(z(u(i-1))) \boxplus z(u(i))\right),
\end{align}
where $i>0$ and $I'(z(u(0)))=I^*(u(0)) \oplus \left(I^*(u(HW-1)) \boxplus z(u(0))\right)$.
Obviously, the keystream used in the diffusion is not private.
Specifically, as for the plain-pixel in position $z(u(i))$, its corresponding random integer used in the modulo addition is actually $z(u(i))$.
\end{itemize}

\subsection{Chosen-plaintext attack on IEACD}

To conceal the relationship between the plain-image and keystream $\mathbf{Y}$, one can generate a pair of plain-images $(\mathbf{I}_0, \mathbf{I}_{1})$, where $\mathbf{I}_0$ is a nature image, $|I_1(a)-I_0(a)|=1$, $I_1(i)=I_0(i)$ for $i\in [0, a)$ and $i\in(a, HW)$, and the index $a$ is any given integer.
Denote the intermediate keystreams and cipher-images corresponding to $(\mathbf{I}_0, \mathbf{I}_{1})$ by $(\mathbf{Y}_0, \mathbf{Y}_{1})$ and $(\mathbf{I}'_0, \mathbf{I}'_{1})$, respectively.
Define the bitwise XOR operation of two plain-images $\mathbf{I}_0$ and $\mathbf{I}_{1}$ as $\mathbf{I}_{0\oplus 1}=\mathbf{I}_0\oplus \mathbf{I}_{1}$, where $\mathbf{I}_0$ and $\mathbf{I}_{1}$ are encrypted by the same secret key.
For simplicity, define a sequence $\mathbf{Z}'=\{z'(i)\}_{i=0}^{HW-1}$, where $z'(i)=z(u(i))$ for $i=0\sim(HW-1)$.

\subsubsection{Determining $\mathbf{Z}$}
\label{ssec:IEACD:recover-z}

As $\mathbf{U}$ is known, one can obtain $\mathbf{Z}$ once $\mathbf{Z}'$ is recovered.
So, one can attempt to determine $\mathbf{Z}'$ first.
Assume $z'(b)=a$ and $b>1$ in the following analysis.
Let $I^*_{0\oplus 1\oplus z'}(u(i-1))$ denote $(I^*_0(u(i-1))\boxplus z'(i))\oplus (I^*_1(u(i-1))\boxplus z'(i))$.
Based on the analysis in Sec.~\ref{sec:IEACD:weak}, one can assume that $\mathbf{Y}_0=\mathbf{Y}_1$.
From Eq.~\eqref{eq:first}, one has
\begin{multline}
 I^*_{0\oplus 1}(u(i))=I^*_{0\oplus 1\oplus z'}(u(i-1))\oplus\\
    \begin{cases}
    0,                    &    \mbox{when $0<i\leq(HW-1)$, $i\neq b$}; \\
  I_{0\oplus 1}(z'(i-1)), & \mbox{when $i=b$},
    \end{cases}
  \label{eq:attack:first1}
\end{multline}
According to Eq.~\eqref{eq:attack:first1}, if $I_0^*(u(i-1))=I_1^*(u(i-1))$ and $i\neq b$, then $I_0^*(u(i))=I_1^*(u(i))$.
From $I^{*}(u(0))=I(z'(0))\oplus Y(u(0))\oplus\left(C \boxplus z'(0)\right)$, one can deduce $I_0^*(u(0))=I_1^*(u(0))$,
and then get $I_0^*(u(i))=I_1^*(u(i))$ for $i=1\sim(b-1)$.
Consequently, Eq.~\eqref{eq:attack:first1} can be represented as
\begin{equation}
\label{eq:attack:first2}
I^*_{0\oplus 1}(u(i))=
\begin{cases}
  0,                    & \mbox{when $0\leq i < b$}; \\
  I_{0\oplus 1}(z'(i)), & \mbox{when $i=b$}; \\
  I^*_{0\oplus 1\oplus z'}(u(i-1)), & \mbox{when $b<i\leq HW-1$}.
\end{cases}
\end{equation}
Referring to Eq.~\eqref{eq:second} and XORing the two cipher-images, one has
\begin{equation*}
I'_{0\oplus 1}(z'(i))=
(I_0'(z'(i-1))\boxplus z'(i))\oplus (I'_1(z'(i-1))\boxplus z'(i))\oplus I_{0\oplus 1}^{*}(u(i)).
\end{equation*}
Then, incorporating Eq.~\eqref{eq:attack:first2} into the above equation, one can get
\begin{subequations}
\begin{flalign*}
\hspace{3mm}
&I'_{0\oplus 1}(z'(i))=(I_0'(z'(i-1))\boxplus z'(i))\oplus(I'_1(z'(i-1))\boxplus z'(i))&
\end{flalign*}\vspace{-5mm}
\begin{numcases}{\oplus}
I_{0\oplus 1}(z'(i)), & when $i=b$;   \label{eq:attack:diffb}\\
0,                    & when $0<i<b$;        \label{eq:attack:diff1}	\\
I^*_{0\oplus 1\oplus z'}(u(i-1)), & when $b<i\leq(HW-1)$, \label{eq:attack:diff2}
\end{numcases}
\end{subequations}
which is the key equation for the attack. As the above equation has three cases, sequence $\mathbf{Z}'$ is likewise divided into three recovery parts:
$z'(b-1)$, $\{z'(i)\}_{i=0}^{b-2}$, and $\{z'(i)\}_{i=b+1}^{HW-1}$, which are discussed separately in the following:
\begin{itemize}
\item Determining $z'(b-1)$

Incorporating $z'(b)=a$ into Eq.~\eqref{eq:attack:diffb}, one has
\begin{equation}
I'_{0\oplus 1}(a)=(I_0'(z'(b-1))\boxplus a)\oplus(I'_1(z'(b-1))\boxplus a)\oplus I_{0\oplus 1}(a).
\label{eq:hold}
\end{equation}
Enumerating $z'(b-1)=j$, one can obtain a set containing all possible values of $z'(b-1)$ via Eq.~(\ref{eq:hold}), where $j=0\sim(HW-1)$ and $j\neq a$.
Adopting more known plain-images and the corresponding cipher-images, one can get more different sets and intersect them, which makes
the probability $z'(b-1)$ is correctly determined approach one. Ideally, the probability is one if and only if the cardinality of the intersection of these sets is equal to one.

\item Determining $\{z'(i)\}_{i=0}^{b-2}$

Substituting $i$ with $b-1$ in Eq.~\eqref{eq:attack:diff1}, one can get
\begin{multline}
I'_{0\oplus 1}(z'(b-1))=(I_0'(z'(b-2))\boxplus z'(b-1))\oplus\\ (I'_1(z'(b-2))\boxplus z'(b-1)).
\label{eq:hold2}
\end{multline}
In the above equation, $z'(b-1)$ is determined and only $z'(b-2)$ is unknown.
Similar to the recovery of $z'(b-1)$, one can enumerate $z'(b-2)=j$ and verify it via Eq.~(\ref{eq:hold2}), where $j\in\{0, 1, \cdots, HW-1\}$ and $j\neq z'(b-1)$.
As every element in $\mathbf{Z}'$ is unique, the derived element should be recorded and not used in the following enumeration.
Apparently, the elements before $z'(b-2)$ in $\mathbf{Z}'$ can be likewise determined via Eq.~\eqref{eq:attack:diff1}.
Since $z'(0)$ is the first element, this process is naturally finished when no element can be found using Eq.~\eqref{eq:attack:diff1}.
After $\{z'(i)\}_{i=0}^{b-2}$ is obtained, $b$ is also determined by the way.
In case of $b=HW-1$, $\mathbf{Z}'$ is completely recovered.
But in the other cases, the elements after $z'(b)$ in $\mathbf{Z}'$ remain undetermined at this moment.

\item Determining $\{z'(i)\}_{i=b+1}^{HW-1}$

Now, determine $z'(b+1)$ via Eq.~\eqref{eq:attack:diff2}.
When $i=b+1$, Eq.~\eqref{eq:attack:diff2} becomes
\begin{multline}
\label{eq:attack:diff2b}
I'_{0\oplus 1}(z'(b+1))=(I_0'(z'(b))\boxplus z'(b+1))\oplus(I'_1(z'(b))\boxplus z'(b+1))\\
\oplus(I^*_0(u(b))\boxplus z'(b+1))\oplus(I^*_1(u(b))\boxplus z'(b+1)).
\end{multline}
Besides $z'(b+1)$, $I^*_0(u(b))$ and $I^*_1(u(b))$ are still unknown in Eq.~\eqref{eq:attack:diff2b}.
Since $z'(b-1)$ is obtained, they can be calculated via
\begin{equation}
  I^{*}(u(b)) = I'(z'(b)) \oplus \left( I'(z'(b-1)) \boxplus z'(b) \right),
  \label{eq:ub}
\end{equation}
which is derived from Eq.~\eqref{eq:second}.
Similarly, $\{I^{*}_0(u(i))\}_{i=1}^{b-1}$ and $\{I^{*}_{1}(u(i))\}_{i=1}^{b-1}$ also can be calculated.
Just as determining $\{z'(i)\}_{i=0}^{b-1}$, $z'(b+1)$ can be confirmed through enumeration and verification via Eq.~\eqref{eq:attack:diff2b}.
Again, one can calculate $I^*_0(u(b+1))$ and $I^*_1(u(b+1))$ via Eq.~(\ref{eq:ub}),
and then determine $z'(b+2)$ by Eq.~\eqref{eq:attack:diff2}.
By this way, the rest can also be determined one by one in turn.
\end{itemize}

In the process of determining $\mathbf{Z}'$, as for a known or given element $z'(i)$, one attempts to find its neighbor by verifying whether the corresponding equation holds.
Therefore, $\mathbf{Z}'$ is reconstructed by seeking the relative positions of elements.
As mentioned before, after constructing $\mathbf{Z}'$, the permutation vector $\mathbf{Z}$ can be restored via $z(u(i))=z'(i)$ for $i=0\sim (HW-1)$.

Now, the attack in case of $b\leq 1$ is discussed.
In fact, the special cases of $b$ can be identified through Eq.~\eqref{eq:attack:diffb} and \eqref{eq:attack:diff1}.
If $b=1$, since $z'(b-1)$ is the first element in $\mathbf{Z}'$, no element can be found via Eq.~\eqref{eq:attack:diff1}.
Hence, one should determine $z'(0)$ via Eq.~\eqref{eq:attack:diffb} and then find the remainder of $\mathbf{Z}'$ through Eq.~\eqref{eq:attack:diff2}.
If $b=0$, the attack is failed. Since Eq.~\eqref{eq:attack:diffb} and \eqref{eq:attack:diff1} both no longer hold, one would directly attempt to determine $z'(1)$ through
\begin{multline*}
I'_{0\oplus 1}(z'(1))=(I_0'(z'(0))\boxplus z'(1))\oplus(I'_1(z'(0))\boxplus z'(1))
\oplus(I^*_0(u(0))\\\boxplus z'(1))\oplus(I^*_1(u(0))\boxplus z'(1)),
\end{multline*}
which is derived from Eq.~\eqref{eq:attack:diff2}.
Here, $I^*_0(u(0))$ and $I^*_1(u(0))$ are still unknown, and only $I^*_{0\oplus 1}(u(0))=I_{0\oplus 1}(z'(0))$ can be obtained from Eq.~\eqref{eq:attack:first2}.
Apparently, the available information is insufficient to obtain $I^*_0(u(0))$ and $I^*_1(u(0))$, and then the attack cannot proceed.
Such case occurs with a low probability $\frac{1}{HW}$, which is $1.52\times 10^{-5}$ when $H=W=256$.
So it does not impact the attack much. If it occurs, one just needs to choose a different index $a'$ and generate the corresponding plain-images again.

Next, let us investigate how many plain-images are sufficient to recover $\mathbf{Z}'$ exactly.
As there is a strong correlation between $\mathbf{I}_0'$ and $\mathbf{I}_1'$, it is intractable to estimate it.
Therefore, assume that each pixel in $M_{\rm p}$ pairs of cipher-images follows independently identical distribution in subsequent discussion.
The retrieve process of $\mathbf{Z}'$ is similar to the attack method on permutation-only ciphers given in \cite{Lcq:Optimal:SP11,cqli:scramble:IEEEM17}, which attempts to find the sole exact permutation position from a set containing all possible positions. As discussed in \cite{Lcq:Optimal:SP11,cqli:scramble:IEEEM17},
some minor error elements in permutation matrix have no much influence on the decryption performance.
However, due to the two-round crossover diffusion, any wrong element in $\mathbf{Z}'$ can incur that the decryption result has no any visual information. In other words, the error-tolerant rate of $\mathbf{Z}'$ for decryption performance is zero.

When one attempts to determine $z'(i-1)$ via Eq.~\eqref{eq:attack:diff1} and $z'(i)$ is known, among 65536 combinations of $I_0'(z'(i-1))$ and $I_1'(z'(i-1))$,
only 256 ones satisfy the equation.
And Eq.~\eqref{eq:attack:diff1} should satisfy for $M_{\rm p}$ pairs of cipher-images.
Thus, the possibility deriving a wrong value as the neighbor of a given element is $\frac{1}{256^{M_{\rm p}}}$.
As for Eq.~\eqref{eq:attack:diffb} and \eqref{eq:attack:diff2}, the analysis is similar and the corresponding possibility is the same.
Then, the probability that an element only has sole exact candidate after enumeration and verification is roughly
\begin{equation*}
  P_{\rm s} = \left(1-\frac{1}{256^{M_{\rm p}}}\right)^{HW-2}.
\end{equation*}
Determining $\mathbf{Z}'$ exactly relies on three conditions: $\mathbf{Y}_0=\mathbf{Y}_1$, $b\neq 0$, and the exact neighbor for each element can be obtained.
Hence, the probability of recovering $\mathbf{Z}'$ exactly can be calculated by $P_z=(1-\frac{1}{HW}) \cdot (1-P_{\rm c})^{M_{\rm p}} \cdot (P_{\rm s})^{HW}$.
When $M_{\rm p}=5$ and $H=W=256$, the probability is about 0.993. So it is expected that $\mathbf{Z}'$ can be always recovered exactly when $M_{\rm p}\geq 5$.
The possibility $P_z$ can also be regarded as the attacking success rate.
Apparently, the time complexity of the whole recovering process is $O(M_{\rm p} \cdot (HW)^2)$.

\subsubsection{Determining $N_0$, $C$, and $Y^{\rm L}$}

Now $\mathbf{Z}$, $\{I^{*}_0(u(i))\}_{i=1}^{HW-1}$, and $\{I^{*}_{1}(u(i))\}_{i=1}^{HW-1}$ are known, only the confusion part is left.
The unknown elements of $\mathbf{I}^*_0$ and $\mathbf{I}^*_1$, $I_0^*(u(0))$ and $I_1^*(u(0))$, can be calculated via
\begin{equation*}
  I^*(u(0)) =I'(z'(0)) \oplus \left(I^*(u(HW-1)) \boxplus z'(0)\right).
\end{equation*}
And the elements of $\mathbf{Y}$ can be recovered through
\begin{equation*}
  Y(u(i))=I^{*}(u(i)) \oplus I(z'(i)) \oplus ( I^{*}(u(i-1)) \boxplus z'(i) )
\end{equation*}
except $Y(0)$ ($u(0)=0$).
Similar to Sec.~\ref{sec:IEATD:kpa}, one can guess $N_0$ through brute-force searching and ignoring the influence of the unrecovered value in comparison (assuming $Y(0)=Y(N(0))$).
Then, one can calculate
\begin{equation*}
  C=\left(I^{*}(u(0))\oplus I(z'(0))\oplus Y(u(0)) \right)\boxminus z'(0),
\end{equation*}
where $a\boxminus b=(a-b+256)\bmod 256$.
To decrypt cipher-images completely, choose a plain-image of fixed value 255 as \cite{lim:improve:IM18}.
Thus, one can calculate the corresponding sequence $\mathbf{Y}_{255}$ and extract the longest vector $Y^{\rm L}_{255}$ from the sequence.
Finally, the equivalent secret key $(N_0, C, \mathbf{Z}, Y^{\rm L}_{255})$ can be obtained.

\begin{figure}[!htb]
  \centering
  \begin{minipage}[t]{\ThreeImW}
    \centering
    \includegraphics[width=\ThreeImW]{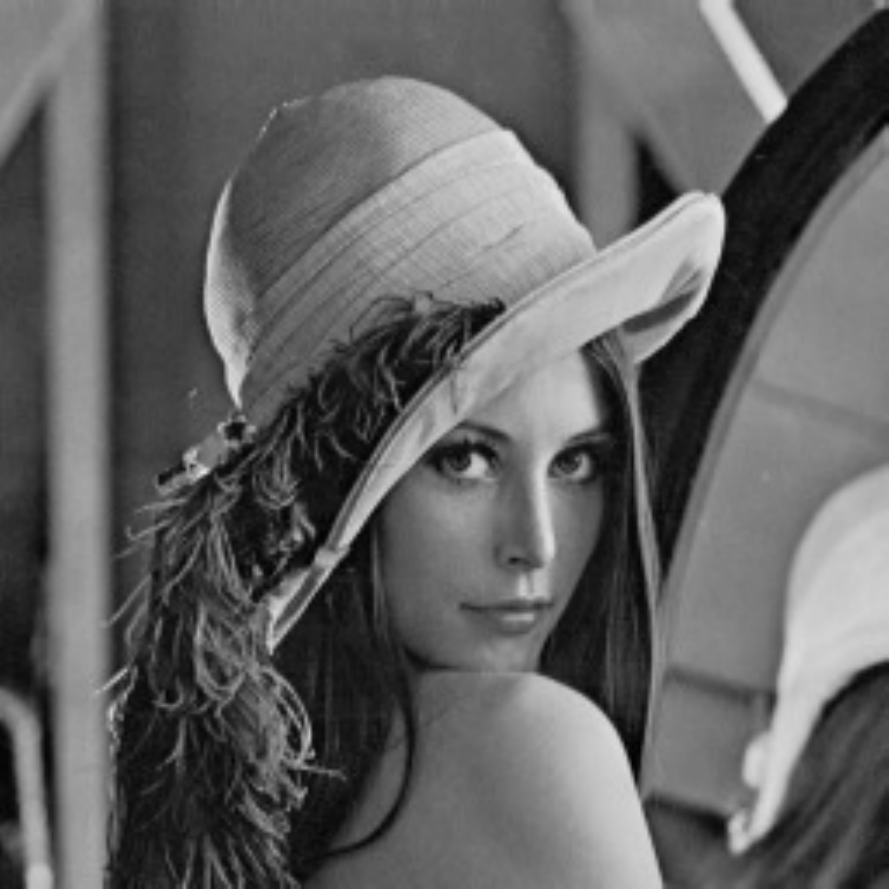}
     $\mathbf{I}_0$
  \end{minipage}
  \begin{minipage}[t]{\ThreeImW}
    \centering
    \includegraphics[width=\ThreeImW]{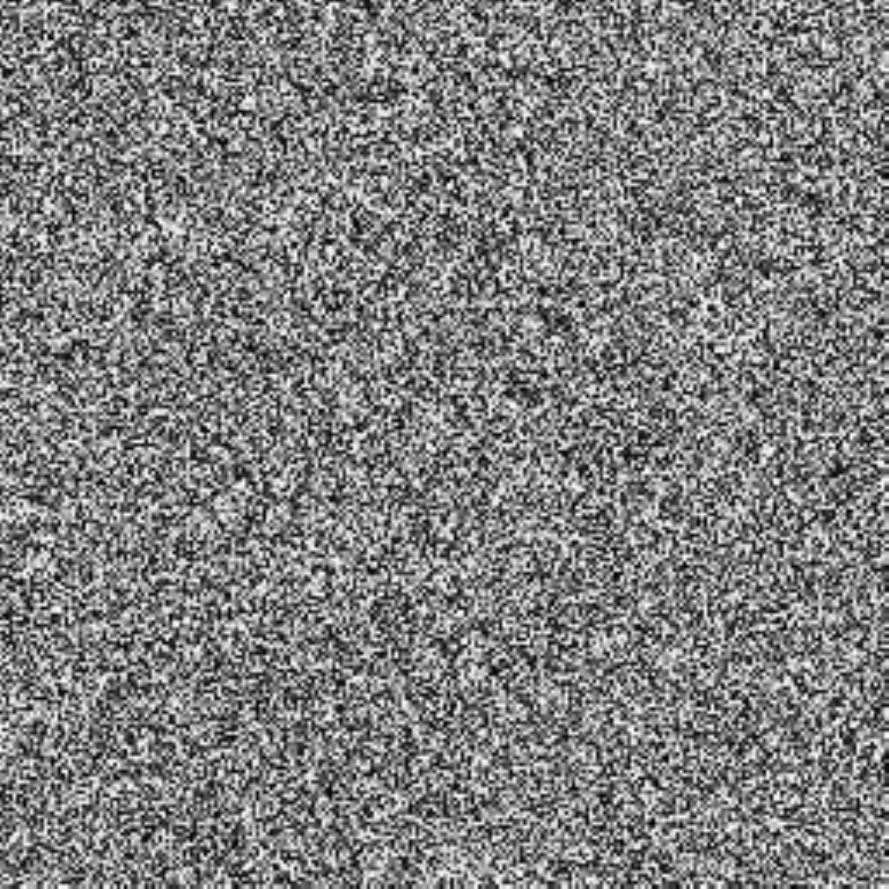}
     $\mathbf{I}_0'$
  \end{minipage}\\
  \begin{minipage}[t]{\ThreeImW}
    \centering
    \includegraphics[width=\ThreeImW]{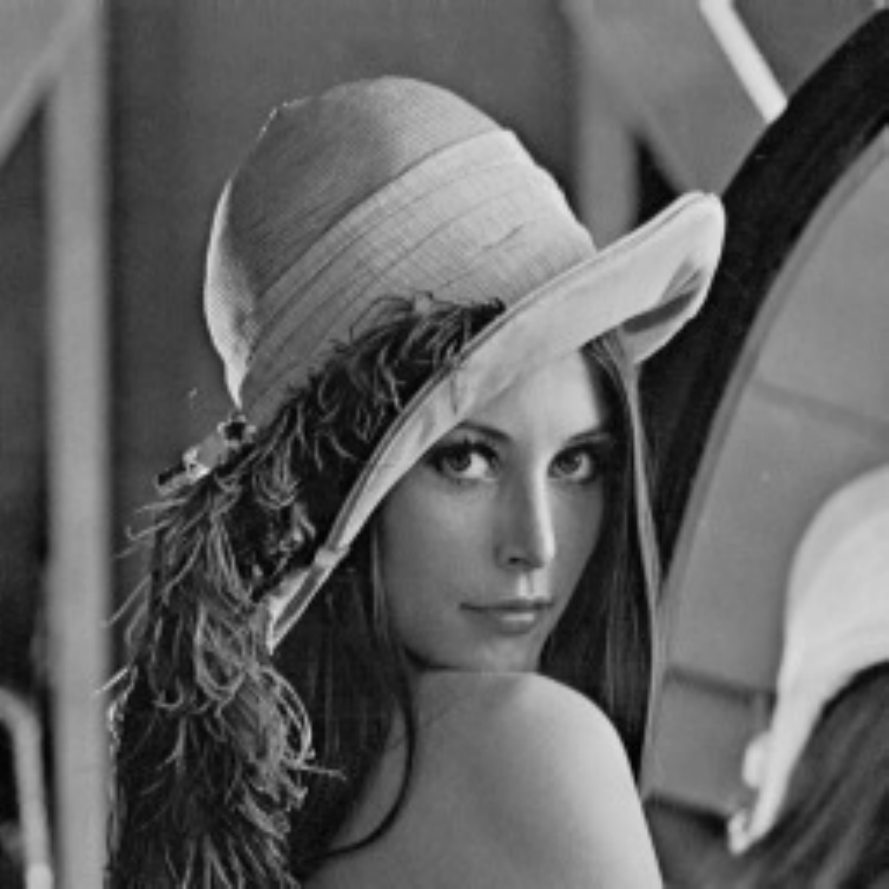}
      $\mathbf{I}_{1}$
  \end{minipage}
  \begin{minipage}[t]{\ThreeImW}
    \centering
    \includegraphics[width=\ThreeImW]{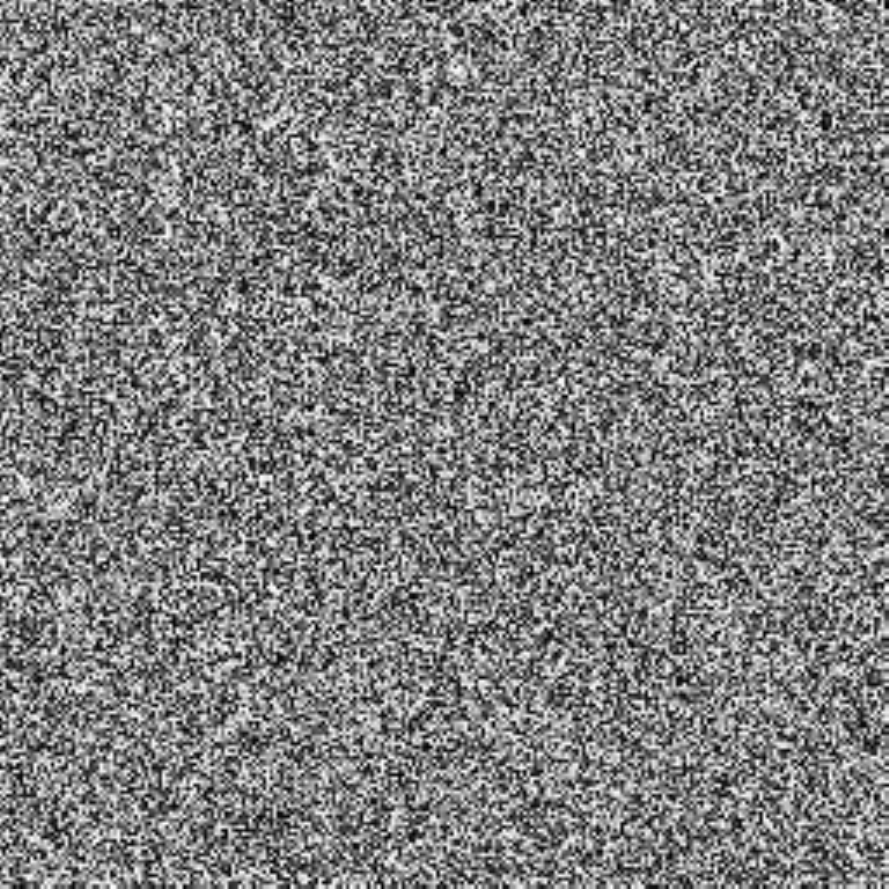}
     $\mathbf{I}_1'$
  \end{minipage}
  \caption{Two similar plain-images and the corresponding cipher-images.}
  \label{fig:cipher}
\end{figure}

\begin{figure}[!htb]
  \centering
  \begin{minipage}{\ThreeImW}
    \centering
    \includegraphics[width=\ThreeImW]{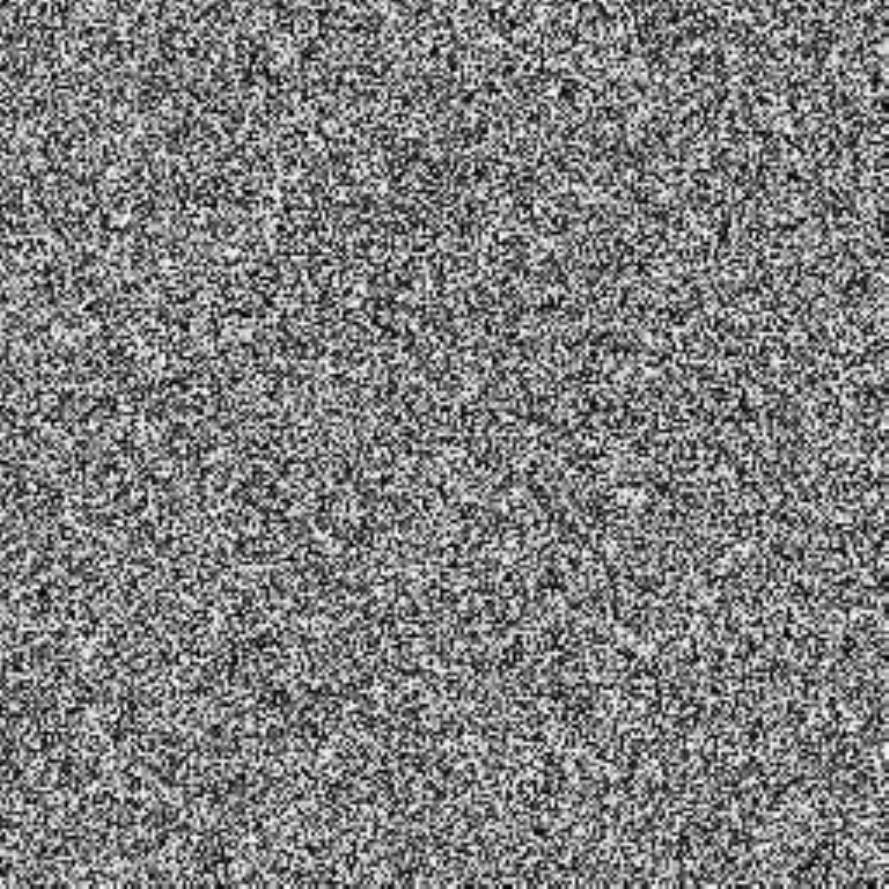}
    a)
  \end{minipage}
  \begin{minipage}{\ThreeImW}
    \centering
    \includegraphics[width=\ThreeImW]{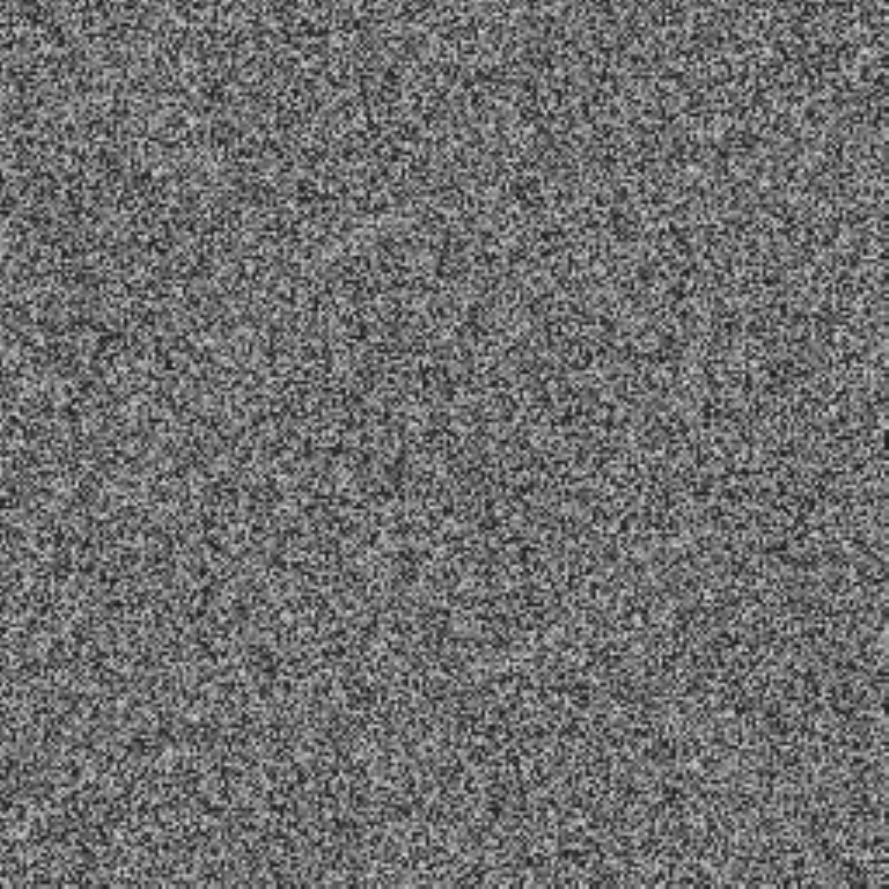}
    b)
  \end{minipage}
  \begin{minipage}{\ThreeImW}
    \centering
    \includegraphics[width=\ThreeImW]{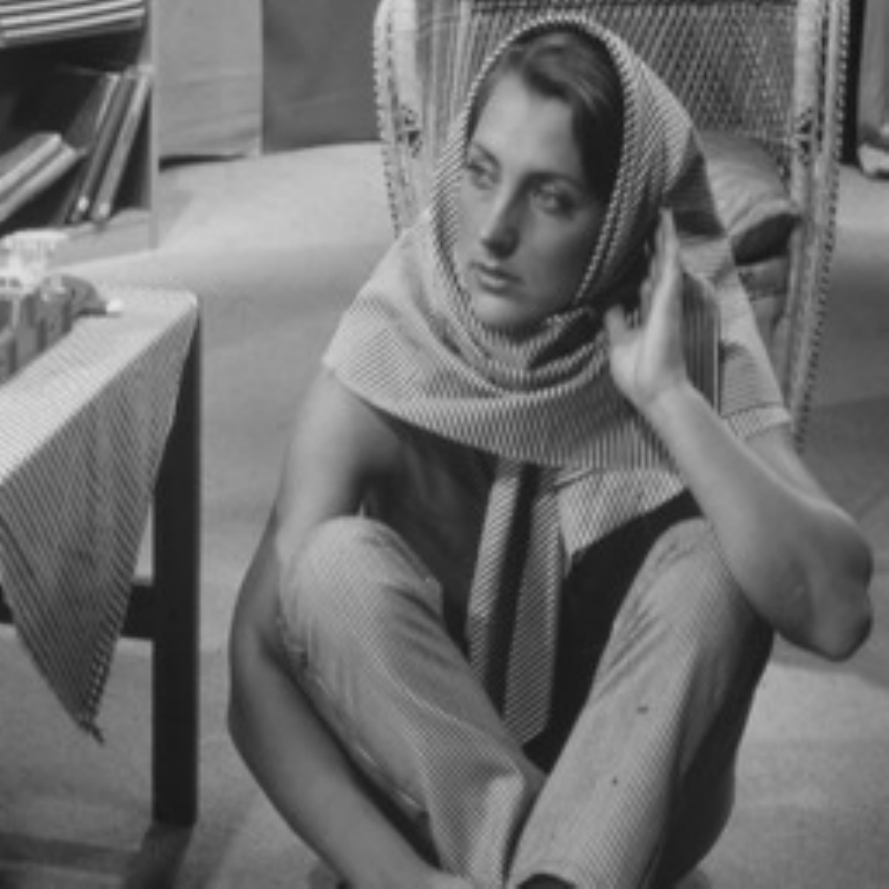}
    c)
  \end{minipage}
  \caption{Two intermediate images and a decrypted image: a) the image removed crossover diffusion; b) the image removed confusion; c) the decrypted result.}
  \label{fig:decipher}
\end{figure}

To test the real performance of the preceding chosen-plaintext attack, some experiments were performed.
As \cite{Mannai:delay:ND15,lim:improve:IM18}, the typical secret key is set as $N_0=100$, $C=216$, $\alpha=6$, $m=19.5$, $h=0.1$, $q(0)=0.75$, and $\beta=3.7488464$.
The initial condition $\mathbf{X}(0)$ of the discretized Ikeda system is a randomly generated vector of length 50.
Figure~\ref{fig:cipher} demonstrates two plain-images used for determining $\mathbf{Z}$ and the corresponding cipher-images, where the index of the changed pixel is 46240.
It is found that the sequence $\mathbf{Z}$ can be recovered with five pairs of plain-images and the corresponding cipher-images.
After $N_0$ and $C$ are determined, $Y^{\rm L}_{255}$ is derived using a plain-image of fixed value 255.
Two corresponding intermediate images and the final result decrypted using the equivalent secret key are shown in Fig.~\ref{fig:decipher}.
To show the attack vividly, Table.~\ref{table:encryption} and \ref{table:attack} list the encryption process of a sample image of size $4\times 4$ and
the corresponding attacking results, respectively.

\setlength\tabcolsep{1pt} 

\begin{table*}[!htb]
  \centering
  \caption{Encryption process of an image of size $4\times 4$ by IEACD.}
  \label{table:encryption}
  \begin{tabular}{*{16}{c}c}
    \toprule
     Item &  1 & 2 & 3 & 4 & 5 & 6 & 7 & 8 & 9 & 10 & 11 & 12 & 13 & 14 & 15 & 16 \\
    \midrule
    $\mathbf{I}_0$ &7&5&7&3&3&10&0&1&6&3&3&2&3&6&5&6 \\
    $\mathbf{Z}$ &14&6&4&12&2&15&10&0&8&7&9&1&11&3&5&13 \\
    $\mathbf{I}_0^{\star}$ & 5&0&3&3&7&6&3&7&6&1&3&5&2&3&10&6 \\
    $\mathbf{N}$ & 4&6&6  \\
    $\mathbf{Y}_0$ &233&33&101&80&233&33&101&80&187&24&233&33&101&80&187&24 \\
    $\mathbf{I}_0^{\star\star}$ &236&33&102&83&238&39&102&87&189&25&234&36&103&83&177&30 \\
    $\mathbf{I}_0^*$ &10&148&224&92&149&241&215&58&175&130&3&121&199&167&109&89 \\
    $\mathbf{I}_0^{**}$ & 109&116&29&109&140&174&247&171&218&249&37&23&80&22&145&225 \\
    $\mathbf{I}'_0$ & 171&23&140&22&29&145&116&249&218&37&247&80&109&225&109&174 \\
    \bottomrule
  \end{tabular}
\end{table*}

\begin{table*}[!htb]
  \centering
  \caption{The items obtained in an attack.}
  \label{table:attack}
  \begin{tabular}{ll}
    \toprule
    Items & The corresponding value  \\
    \midrule
    $a$ & 7 \\
    $z'(b-1)$ & 6  \\
    $\{z'(i)\}_{i=0}^{b-2}$ & $14,8$ \\
    $b$ & 3 \\
    $\{z'(i)\}_{i=b+1}^{HW-1}$ & $4,9,12,1,2,11,15,3,10,5,0,13$ \\
    $\mathbf{I}_0^*$ & $10,148,227,65,134,34,230,13,175,129,6,102,246,118,90,4$ \\
    $\{Y_0(i)\}_{i=1}^{HW-1}$ & $33,101,80,233,33,101,80,187,24,233,33,101,80,187,24$ \\
    $N_0$ & 4 \\
    $C$ & 216 \\
    $Y^{\rm L}$ & $233,33,101,80,187,24,172,246,41,29,12,24$ \\
    \bottomrule
  \end{tabular}
\end{table*}

\section{Conclusion}

This paper analyzed security performance of an image encryption algorithm based on a first-order time-delay system IEATD and the enhanced version IEACD.
Although another research group proposed a chosen-plaintext attack on IEATD, we presented an enhanced attack using the correlation between adjacent vectors of one plain-image and the corresponding cipher-image. Although IEACD is designed by the attacking group with intention to fix the security defects of IEATD, there still exist
some security pitfalls, such as invalid secret keys, insensibility of keystream generation mechanics, and improper configuration of keystream.
Based on these, we designed an efficient chosen-plaintext attack and verified it with extensive experiments.
The serious insecurity of the two algorithms cannot be improved by simple modifications. They can work as typical counterexamples
to remind us to recast scenario-oriented image encryption algorithms following the guidelines and lessons summarized in
\cite{Shannon:CommunicationTheory1949,Preishuber:motivation:TIFS2018,cqli:meet:JISA19}.


\section*{Acknowledgements}

This work was supported by the National Natural Science Foundation of China (no.~61772447),
Scientific Research Fund of Hunan Provincial Education Department (no. 20C1759),
and Science and Technology Program of Changsha (no. kq2004021).

\bibliographystyle{IEEEtran}
\bibliography{IEFTS}

\begin{thebibliography}{10}
\providecommand{\url}[1]{#1}
\csname url@samestyle\endcsname
\providecommand{\newblock}{\relax}
\providecommand{\bibinfo}[2]{#2}
\providecommand{\BIBentrySTDinterwordspacing}{\spaceskip=0pt\relax}
\providecommand{\BIBentryALTinterwordstretchfactor}{4}
\providecommand{\BIBentryALTinterwordspacing}{\spaceskip=\fontdimen2\font plus
\BIBentryALTinterwordstretchfactor\fontdimen3\font minus
  \fontdimen4\font\relax}
\providecommand{\BIBforeignlanguage}[2]{{%
\expandafter\ifx\csname l@#1\endcsname\relax
\typeout{** WARNING: IEEEtran.bst: No hyphenation pattern has been}%
\typeout{** loaded for the language `#1'. Using the pattern for}%
\typeout{** the default language instead.}%
\else
\language=\csname l@#1\endcsname
\fi
#2}}
\providecommand{\BIBdecl}{\relax}
\BIBdecl

\bibitem{abu:CA:IM12}
A.~L. Abu~Dalhoum, B.~A. Mahafzah, A.~A. Awwad, I.~Aldamari, A.~Ortega, and
  M.~Alfonseca, ``Digital image scrambling using {2D} cellular automata,''
  \emph{IEEE Multimedia}, vol.~19, no.~4, pp. 28--36, 2012.

\bibitem{yegd:auto:IM16}
G.~Ye and X.~Huang, ``An image encryption algorithm based on autoblocking and
  electrocardiography,'' \emph{IEEE Multimedia}, vol.~23, no.~2, pp. 64--71,
  2016.

\bibitem{cqli:meet:JISA19}
C.~Li, Y.~Zhang, and E.~Y. Xie, ``When an attacker meets a cipher-image in
  2018: a year in review,'' \emph{Journal of Information Security and
  Applications}, vol.~48, p. art. no. 102361, 2019.

\bibitem{Mannai:delay:ND15}
O.~Mannai, R.~Bechikh, H.~Hermassi, R.~Rhouma, and S.~Belghith, ``A new image
  encryption scheme based on a simple first-order time-delay system with
  appropriate nonlinearity,'' \emph{Nonlinear Dynamics}, vol.~82, pp. 107--117,
  2015.

\bibitem{Jolfaei:Permut:TIFS16}
A.~Jolfaei, X.-W. Wu, and V.~Muthukkumarasamy, ``On the security of
  permutation-only image encryption schemes,'' \emph{IEEE Transactions on
  Information Forensics and Security}, vol.~11, no.~2, pp. 235--246, 2016.

\bibitem{cqli:autoblock:IEEEM18}
C.~Li, D.~Lin, J.~L\"u, and F.~Hao, ``Cryptanalyzing an image encryption
  algorithm based on autoblocking and electrocardiography,'' \emph{IEEE
  Multimedia}, vol.~25, no.~4, pp. 46--56, 2018.

\bibitem{Preishuber:motivation:TIFS2018}
M.~Preishuber, T.~Huetter, S.~Katzenbeisser, and A.~Uhl, ``Depreciating
  motivation and empirical security analysis of chaos-based image and video
  encryption,'' \emph{IEEE Transactions on Information Forensics and Security},
  vol.~13, no.~9, pp. 2137--2150, 2018.

\bibitem{Chenjx:psn:TCSVT21}
J.~Chen, L.~Chen, and Y.~Zhou, ``Cryptanalysis of image ciphers with
  permutation-substitution network and chaos,'' \emph{IEEE Transactions on
  Circuits and Systems for Video Technology}, vol.~31, no.~6, pp. 2494--2508,
  2021.

\bibitem{Shannon:CommunicationTheory1949}
C.~E. Shannon, ``Communication theory of secrecy systems,'' \emph{Bell System
  Technical Journal}, vol.~28, no.~4, pp. 656--715, 1949.

\bibitem{chai:DNA:SP19}
X.~Chai, X.~Fu, Z.~Gan, Y.~Lu, and Y.~Chen, ``A color image cryptosystem based
  on dynamic dna encryption and chaos,'' \emph{Signal Processing}, vol. 155,
  pp. 44--62, 2019.

\bibitem{Hua2021ND}
Z.~Hua, Z.~Zhu, Y.~Chen, and Y.~Li, ``Color image encryption using orthogonal
  latin squares and a new 2{D} chaotic system,'' \emph{Nonlinear Dynamics},
  vol. 104, p. 4505–4522, 2021.

\bibitem{ikeda:1980:optical}
K.~Ikeda, H.~Daido, and O.~Akimoto, ``Optical turbulence: chaotic behavior of
  transmitted light from a ring cavity,'' \emph{Physical Review Letters},
  vol.~45, no.~9, p. 709, 1980.

\bibitem{lim:improve:IM18}
M.~Li, H.~Fan, Y.~Xiang, Y.~Li, and Y.~Zhang, ``Cryptanalysis and improvement
  of a chaotic image encryption by first-order time-delay system,'' \emph{IEEE
  Multimedia}, vol.~25, no.~3, pp. 92--101, 2018.

\bibitem{Ding:period:ND19}
C.~Fan and Q.~Ding, ``Analysing the dynamics of digital chaotic maps via a new
  period search algorithm,'' \emph{Nonlinear Dynamics}, vol.~97, no.~1, pp.
  831--841, 2019.

\bibitem{cqli:network:TCASI2019}
C.~Li, B.~Feng, S.~Li, J.~Kurths, and G.~Chen, ``Dynamic analysis of digital
  chaotic maps via state-mapping networks,'' \emph{IEEE Transactions on
  Circuits and Systems I: Regular Papers}, vol.~66, no.~6, pp. 2322--2335,
  2019.

\bibitem{cqli:Cat:TC21}
C.~Li, K.~Tan, B.~Feng, and J.~L\"u, ``The graph structure of the generalized
  discrete arnold cat map,'' \emph{IEEE Transactions on Computers}, 2021.

\bibitem{Lcq:Optimal:SP11}
C.~Li and K.-T. Lo, ``Optimal quantitative cryptanalysis of permutation-only
  multimedia ciphers against plaintext attacks,'' \emph{Signal Processing},
  vol.~91, no.~4, pp. 949--954, 2011.

\bibitem{cqli:scramble:IEEEM17}
C.~Li, D.~Lin, and J.~Lu, ``Cryptanalyzing an image-scrambling encryption
  algorithm of pixel bits,'' \emph{IEEE Multimedia}, vol.~24, no.~3, pp.
  64--71, 2017.

\end{thebibliography}

\subsection*{  } %

\noindent \textbf{Sheng Liu} is a graduate student in computer science at the School of Computer Science and Electronic Engineering, Hunan University.
His research interests include image privacy protection and image forensics. Contact him at shengliu@hnu.edu.cn.\\

\noindent \textbf{Chengqing Li} is a professor with the College of Information Engineering, Xiangtan University, China.
His research interests include image privacy protection and multimedia cryptanalysis. Li received a PhD in electronic engineering from City University of Hong Kong.
He is the corresponding author of this article. Contact him at chengqingg@gmail.com. \\

\noindent \textbf{Qiao Hu} is an assistant professor with the College of Computer Science and Electronic Engineering, Hunan University, China.
His research interests include RFID security and privacy, cloud computing, and wireless communication security.
Hu received a PhD in information security from City University of Hong Kong. Contact him at huqiao@hnu.edu.cn.

\end{document}